\documentclass[twocolumn,prl,amsmath,amssymb,showpacs,superscriptaddress,floatfix]{revtex4}
\usepackage{graphicx}
\usepackage{bm}

\usepackage{epsf}
\begin{document}
\title{Phase diagram and melting scenarios of two-dimensional Hertzian spheres}

\author{Yu. D. Fomin, E. A. Gaiduk, E. N. Tsiok, and V. N. Ryzhov}
\affiliation{Institute for High Pressure Physics RAS, 108840
Kaluzhskoe shosse, 14, Troitsk, Moscow, Russia}

\date{\today}

\begin{abstract}
We present computer simulations of a system of purely repulsive
soft colloidal particles interacting via the Hertz potential and
constrained to a two-dimensional plane. This potential describes
the elastic interaction of weakly deformable bodies and can be a
reliable model for qualitative description of behavior of soft
macromolecules, like globular micelles and star polymers. We find
a large number of ordered phases, including the dodecagonal
quasicrystal, and analyze the melting scenarios of low density
triangle and square phases. It is interesting that depending on
the position on the phase diagram the system can melt both through
the first order transition and in accordance with the
Berezinskii-Kosterlitz-Thouless-Halperin-Nelson-Young (BKTHNY)
scenario (two continuous transitions with the intermediate hexatic
phase) and also in accordance with recently proposed two-stage
melting with the first order hexatic-isotropic liquid transition
and continuous solid-hexatic transition. We also demonstrate the
possibility of the tricritical point on the melting line.
\end{abstract}


\maketitle

\section{Introduction}

Phase transitions in two-dimensional ($2D$) systems are of great
interest for many reasons. This is a field where usual intuition
can lead to wrong conclusions. Firstly, as it was shown by Landau
and Peierls \cite{lan,p1,p2} there is no two-dimensional crystals
as we understand the crystals in three dimensional ($3D$) space.
Two-dimensional crystal structures do not demonstrate long-range
translational order. However, there is still long-range
orientational order and quasi-long range translational order in
$2D$ crystals \cite{mermin,pu2017}. As a result, the melting of
$2D$ crystalline phases can occur not only as the first order
phase transition \cite{chui83,ryzhovJETP,RT1,rto1,rto2}, like it
always happens in $3D$, but a number of different scenarios are
possible . In addition to the usual first order phase transition
melting of the $2D$ crystal can appear as two continuous
transitions of Berezinskii-Kosterlitz-Thouless (BKT) type
\cite{ber1,ber2,kosthoul72,kosthoul73,kost}
(Berezinskii-Kosterlitz-Thouless-Halpering-Nelson-Young (BKTHNY)
scenario
\cite{pu2017,kosthoul72,kosthoul73,kost,halpnel1,halpnel2,halpnel3,keim1,zanh,keim2,keim3,keim4}).
The BKTHNY scenario seems most popular now. According to this
theory $2D$ solids melt through dissociation of bound dislocation
pairs. As a result, the long-ranged orientational order transforms
into quasi-long-ranged order, and the quasi-long-ranged positional
order becomes short-ranged. The new intermediate phase with
quasi-long-range orientational order is called a hexatic phase (if
melting of square crystal is considered then it is a tetratic
phase). It has a zero shear rigidity and because of this it should
be considered as a kind of ordered liquid. The hexatic phase
transforms into an isotropic liquid phase having short-ranged
orientational and positional orders through unbinding dislocations
pairs. It should be noted, that the BKTHNY theory provides only
limits of stability of the solid and hexatic phases.

Recently, the other melting scenario was proposed
\cite{foh1,foh2,foh3,foh4,foh5,foh6}. In contrast to the BKTHNY
theory, it was shown that the system can melt through  a
continuous BKT type solid-hexatic transition, but the
hexatic-liquid transition is of first-order \cite{foh1,foh2,foh3}.
In paper \cite{foh4} Kapfer and Krauth explored behavior of a soft
disk system with potential  $U(r)=(\sigma/r)^n$. The system was
shown to melt in accordance with the BKTHNY theory for $n\leq 6$,
while for $n>6$ the two-stage melting transition took place with
continuous solid-hexatic and the first-order hexatic-liquid
transition.

Until recently it only triangular lattices were observed in
two-dimensional systems in experiments and computer simulations.
However, in the last decades there appeared a lot of experimental
(for example, graphene, square ice \cite{si}, square iron
\cite{siron}, etc.) and obtained from computer simulations
nontrivial $2D$ ordered structures, including triangular, square,
honeycomb, Kagome, snub-square tiling, quasicrystals etc.
\cite{str1,str2,str3,str4,str5,str6,str7,str8,str9}. This kind of
behavior is characteristic for the $2D$ core-softened potentials
\cite{dfrt1,dfrt2,dfrt3,dfrt4,dfrt5,dfrt6,dfrt7,trus}. The
different forms of core-softened potentials are widely used for
qualitative description of the anomalous water-like behavior,
including density, structural and diffusion anomalies,
liquid-liquid phase transitions, glass transitions, and melting
maxima
\cite{jcp2008,wepre,we_inv,we2011,RCR,we2013-2,buld2009,fr1,fr2,fr3,fr4,barbosa,barbosa1,barbosa2,barbosa3,buld2d,scala,prest2,prest1}.

Here we focus on the bounded soft potential - the Hertzian
spheres. The Hertz potential describes the potential energy of two
elastic spheres under an axial compression \cite{lanl}. The
potential can be written in the following form:
\begin{equation}
   U(r)=\varepsilon \left ( 1- r/ \sigma \right)^{\alpha}.
   \label{h}
\end{equation}
The parameters $\varepsilon$ and $\sigma$ define the energy and
length scales and the exponent $\alpha$ determines the steepness
of the potential \footnote{In some publications the potential is
defined as $U(r)=\frac{\varepsilon}{\alpha} \left ( 1- r/ \sigma
\right)^{\alpha}$. Although it does not affect the qualitative
behavior of the system it leads to different energy scale and
therefore in order to compare the results one needs to recalculate
the temperatures}. Strictly speaking, the potential (\ref{h})
corresponds to slightly deformed elastic spheres and cannot be
applied for very large compressions and densities. However, Hertz
potential may be used as a model potential for soft particles that
is bounded and has finite range. This potential can be used for
qualitative description of behavior of soft uncharged
macromolecules, like globular micelles and star polymers.

The phase diagram of $3D$ Hertz system with $\alpha =5/2$ was
reported in the seminal work by Daan Frenkel and coauthors
\cite{hertz3d}. It was shown that this phase diagram includes
numerous crystalline phases including non-cubic ones. Moreover, it
was found that the fluid phase of the system demonstrates
diffusion anomaly, i.e. the diffusion coefficient increases with
density in some region of densities and temperatures. Later it was
shown that this system also demonstrates structural anomaly
\cite{rosbreak} identified by excess entropy criterion
\cite{strcriteria} and that this system does not obey the
Rosenfeld scaling \cite{ros,ros1} of the diffusion coefficient
\cite{rosbreak}.

The phase diagram of the $2D$ Hertz system was firstly reported in
\cite{miller}. Three values of $\alpha$ were considered: $3/2$,
$5/2$ and $7/2$ and it was shown that the steepness of the
potential crucially changes the phase diagram. In the particular
case of  $\alpha=5/2$ the authors of Ref. \cite{miller} found
several triangular and square crystals and a phase of pentagons.
Although the authors have observed the phase of pentagons they do
not recognize the symmetry of this structure. Moreover, the work
\cite{miller} gives only rough preliminary investigation of the
phase diagram without strict determination of the phase boundaries
and melting scenarios.

The investigation of the phase diagram of Hertz system with
different $\alpha$ was continued in Ref. \cite{hertzqc}. This
paper presents an investigation of phases which appear in the
Hertz system with $\alpha$ varying from $2$ to $3$ at fixed
temperature. Importantly, the sequence of the phases reported for
$\alpha=5/2$ is different from the one in Ref. \cite{miller}. In
Ref. \cite{hertzqc} the authors recognize the phase of pentagons
as dodecagonal quasicrystal (DDQC). They also find that DDQC
transforms into stripe phase. At high density the system
demonstrates rhombohedral phase. Both stripe and rhombohedral
phases were not observed in \cite{miller}.

Another important observation of Ref. \cite{hertzqc} which is in
agreement with Ref. \cite{miller} is that the larger values of
$\alpha$ corresponds to the simpler phase diagrams. Complex
crystalline or quasicrystalline structures appear only at low
$\alpha$ and as $\alpha$ increases the number of different phases
in the system decreases.

The melting scenario of the low density triangular phase of Hertz
disks with $\alpha=5/2$ was reported in Ref. \cite{hertzmelt}. It
was found that this phase demonstrates reentrant melting. The
melting scenario of the low density branch (left branch)
corresponds to the scenario proposed in Refs.
\cite{foh1,foh2,foh3,foh4,foh5,foh6}, i.e. continuous transition
from solid to hexatic phase and the first order one from hexatic
to liquid. However, the melting scenario changes in the point of
maximum of the melting line and melting of the high density branch
(right) branch proceeds in accordance to the BKTHNY scenario, i.e.
two continuous transitions. However, only the limits of stability
of the liquid are shown in Ref. \cite{hertzmelt} (see Fig. 1 of
this publication). In order to make the phase diagram in the
vicinity of the low density triangular phase complete one needs to
add also the lines of transitions between the hexatic phase and
the crystal.

Summing up the results from literature on the phase diagram of
Hertzian disks with $\alpha=5/2$ one can see that they are rather
sporadic. Up to now there is no full phase diagram of the system.
Also the melting scenario is determined only for the triangular
phase with low density. At the same time not all lines of the
transitions at the melting of triangular and square phases are
reported.

The goal of the present paper is to perform a detailed study of
the phase diagram of the Hertz disks with $\alpha=5/2$. We find
the regions of stability of different ordered phases and discuss
the melting scenarios of the low density triangular and square
phases. We will also discuss the mechanism of transformation of
the low density triangular phase into low density square phase.

\section{Systems and methods}

In the present paper we study the Hertz disk system with
$\alpha=5/2$. We use the reduced unit to represent all quantities
in the study, i.e. we use $\varepsilon$ and $\sigma$ as the units
of energy and length correspondingly. Following Ref. \cite{miller}
we consider the densities from $\rho_{min}=0.6$ up to
$\rho_{max}=10$. It is extremely wide region of densities. It is
especially important since the density jumps in two-dimensional
transitions is usually rather low - of the order of $10^{-3} -
10^{-2}$, therefore the regions of phase transitions are very
narrow comparing to the total interval of the densities under
consideration. Because of this we do the simulations in two steps.

At the first step we do rough estimation of the phase diagram. For
doing this we simulate a small system of $4000$ particles at
densities from $\rho_{min}$ up to $\rho_{max}$ with the step
$\Delta \rho =0.2$. If we see that the system demonstrates
different phases at some step in density we decrease the step and
investigate this region in more details.

The system was simulated in a rectangular box. The initial
configuration was set up to be a high temperature liquid
configuration. Starting from this configuration the system was
simulated by means of molecular dynamics method \cite{book_fs}
(LAMMPS package \cite{lammps}) in canonical ensemble (constant
number of particles N, area S and temperature T) for $5 \cdot
10^7$ steps with time step $dt=0.001$. The first $2 \cdot 10^7$
were used for equilibration while during the last $3 \cdot 10^7$
sampling of the system was performed.

In order to characterize the thermodynamic properties of the
system we studied equations of state (eos), i.e. pressure as a
function of the density along isotherms. If a first order phase
transition takes place in the system one observes a Mayer-Wood
loop. The transition points can be found by Maxwell construction.
In case of the continuous transition the eos demonstrates a bend.
Therefore, independently of the nature of the transition eos can
be used as the first signal.

Having identified the regions of phase transitions from the eos we
perform identification of the phases. For doing this we apply
several methods. The most robust method involves the radial
distribution functions $g(r)$. This method allows to distinguish
different crystalline phases. However, in the case of
two-dimensional systems it is not so efficient in distinguishing
liquids and crystals, because the two-dimensional liquids usually
demonstrate rather ordered behavior next to the melting line.

In order to investigate the melting of triangular and square
crystals we used the orientational and translational order
parameters.

The orientational order parameter (OOP) of triangular lattice is
defined as \cite{halpnel1,halpnel2,dfrt5,dfrt6}
\begin{equation}
\Psi_6({\bf r_i})=\frac{1}{n(i)}\sum_{j=1}^{n(i)} e^{i
n\theta_{ij}}\label{psi6loc},
\end{equation}
where $\theta_{ij}$ is the angle of the vector between particles
$i$ and $j$ with respect to a reference axis and the sum over $j$
is counting $n(i)$ nearest-neighbors of $j$, obtained from the
Voronoi construction. The global OOP can be calculated as an
average over all particles:
\begin{equation}
\psi_6=\frac{1}{N}\left<\left|\sum_i \Psi_6({\bf
r}_i)\right|\right>.\label{psi6}
\end{equation}

The square lattice orientational order $\Psi_4({\bf r_i})$ (and
the global one $\psi_4$) is defined in the same way.

In accordance with the standard definitions \cite{halpnel1,
halpnel2, dfrt5, dfrt6}, translational order parameter (TOP) has
the form:
\begin{equation}
\psi_T=\frac{1}{N}\left<\left|\sum_i e^{i{\bf G
r}_i}\right|\right>, \label{psit}
\end{equation}
where ${\bf r}_i$ is the position vector of particle $i$ and {\bf
G} is the reciprocal-lattice vector of the first shell of the
crystal lattice.

We also calculate the correlation functions of OOP and TOP which
contain information on the long-range or quasi-long-range ordering
in the system. The orientational correlation function (OCF)
$G_6(r)$ is given by the following expression:
\begin{equation}
G_6(r)=\frac{\left<\Psi_6({\bf r})\Psi_6^*({\bf 0})\right>}{g(r)},
\label{g6}
\end{equation}
where $g(r)=<\delta({\bf r}_i)\delta({\bf r}_j)>$  is the pair
distribution function. In the hexatic phase the long range
behavior of $G_6(r)$ has the form $G_6(r)\propto r^{-\eta_6}$ with
$\eta_6 \leq \frac{1}{4}$ \cite{halpnel1, halpnel2}.

In case of the triangular crystal there is long range
orientational order and quasi-long-range translational order. It
leads to flat shape of the orientational OCF, whilst OCF decays
algebraically in the hexatic phase and exponentially in the
isotropic liquid.

The translational correlation function (TCF) can be calculated as
\begin{equation}
G_T(r)=\frac{<\exp(i{\bf G}({\bf r}_i-{\bf r}_j))>}{g(r)},
\label{GT}
\end{equation}
where $r=|{\bf r}_i-{\bf r}_j|$. In the solid phase the long range
behavior of $G_T(r)$ has the form $G_T(r)\propto r^{-\eta_T}$ with
$\eta_T \leq \frac{1}{3}$ \cite{halpnel1, halpnel2}. In the
hexatic phase and isotropic liquid $G_T$ decays exponentially.

Analogous correlation functions can be introduced for the square
crystal by substituting $\Psi_6$ by $\Psi_4$ in the case of OCF
and using the correlation function of $\Psi_T$ for the square
crystal $G_T^{sq}$. Unfortunately, there are no calculations of
the exponents $\eta$  at which the square crystal transforms into
tetratic phase and the tetratic phase into isotropic liquid. By
analogy with the triangular crystal one can use $\eta_4=1/4$ and
$\eta_T=1/3$, but these values should be considered as the first
guess rather then exact ones.

An important shortcoming of the order parameters introduced above
is that they are applicable to a single crystal structure only.
That is why if one finds a novel crystal structure with different
symmetry the order parameters can be zero in this structure. That
is why it can be difficult to find out the symmetry of the
structure using the order parameters only. This problem can be
solved by calculating the diffraction patterns. The diffraction
patterns are the intensity maps of the static structure factor:

\begin{equation}
  S({\bf k})=\frac{1}{N} < \left( \sum_{i=1}^N cos^2({\bf kr}_i)
  \right)^2+\left( \sum_{i=1}^N sin^2({\bf kr}_i) \right)^2>.
\end{equation}

Combining the information obtained from radial distribution
functions, orientational and translational order parameters and
diffraction patterns we are able to identify the structures
observed in the system.

\section{Results and discussion}

\begin{figure}
\begin{center}
\includegraphics[width=4cm,height=4cm]{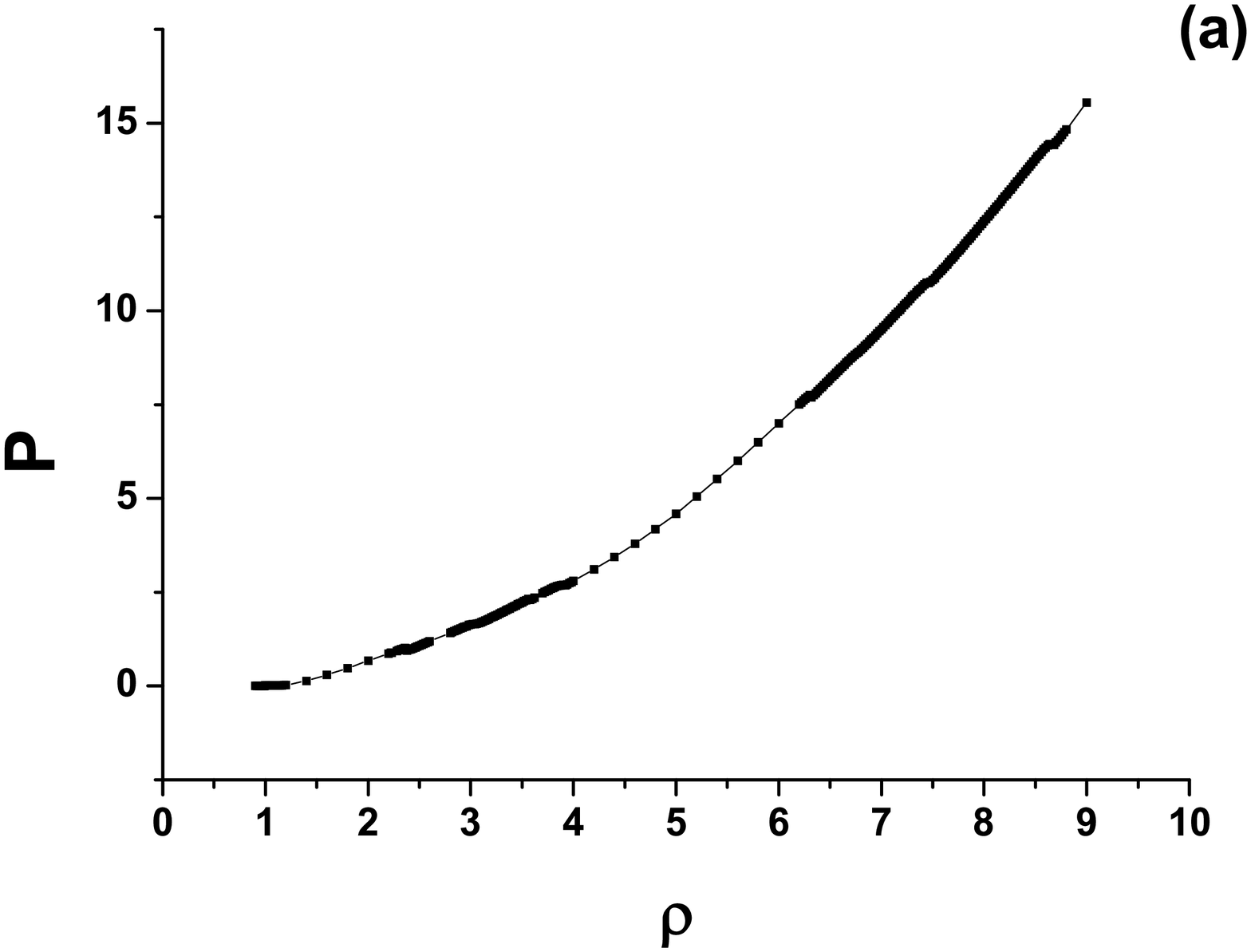}%
\includegraphics[width=4cm,height=4cm]{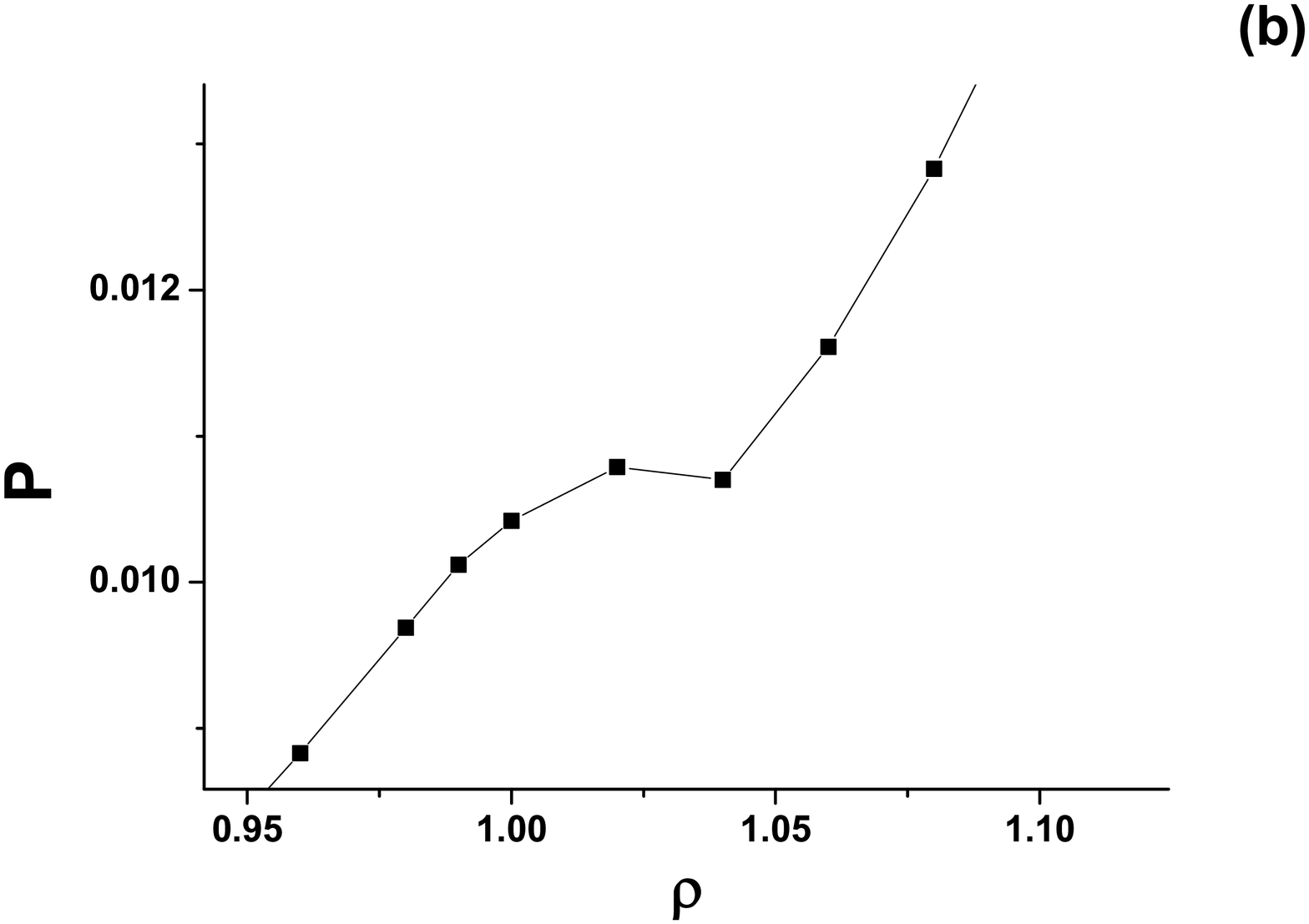}%

\includegraphics[width=4cm,height=4cm]{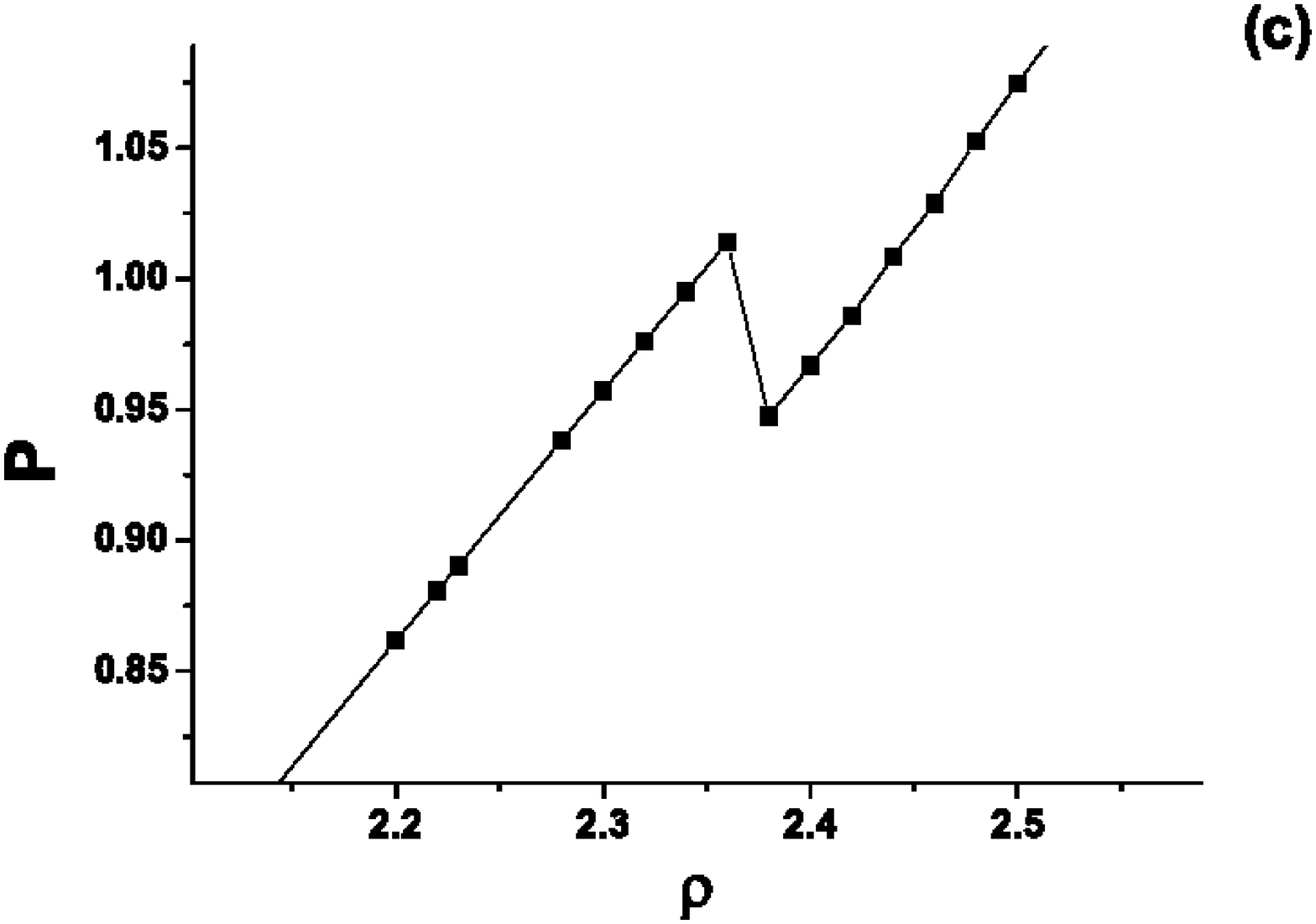}%
\includegraphics[width=4cm,height=4cm]{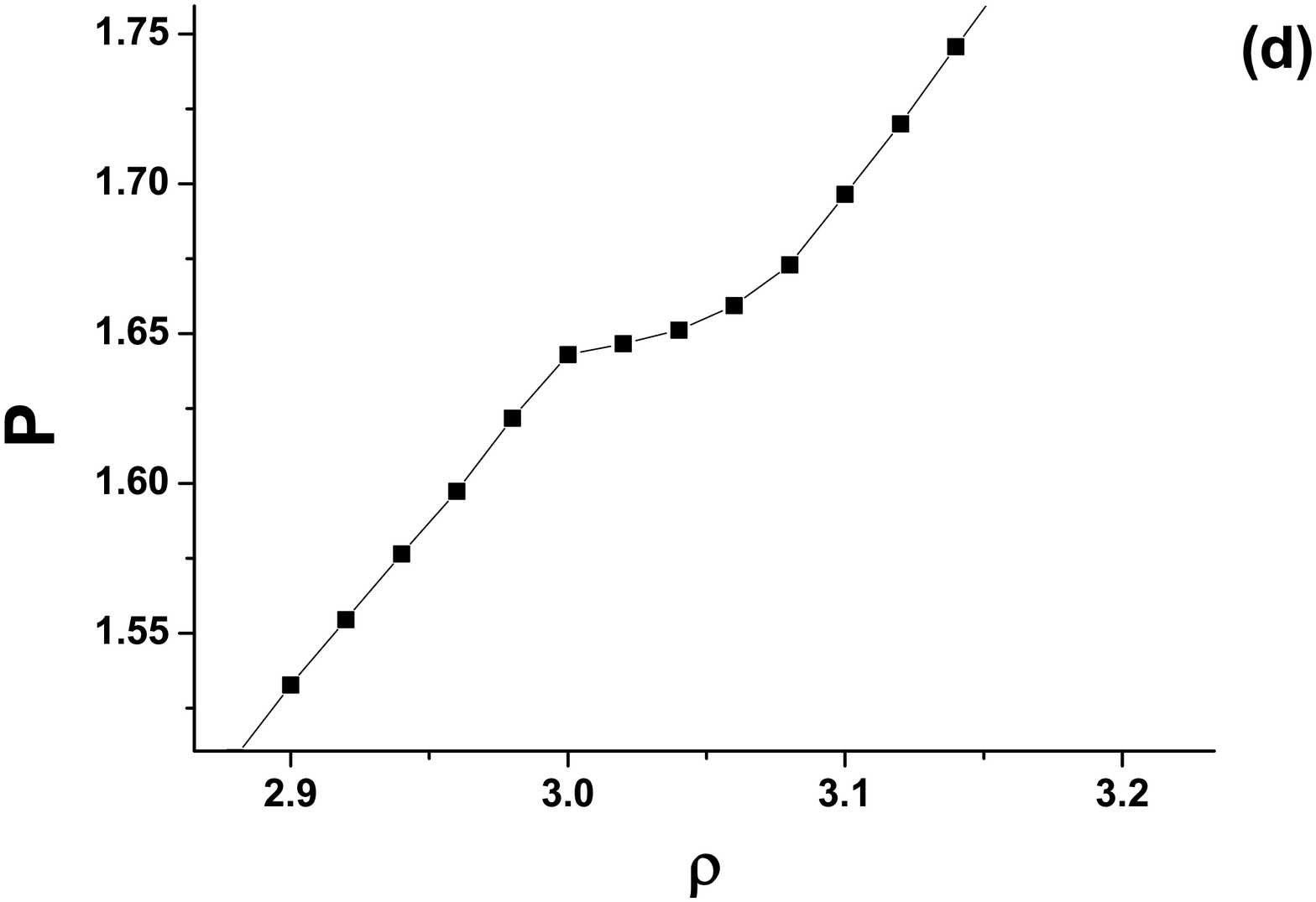}%

\includegraphics[width=4cm,height=4cm]{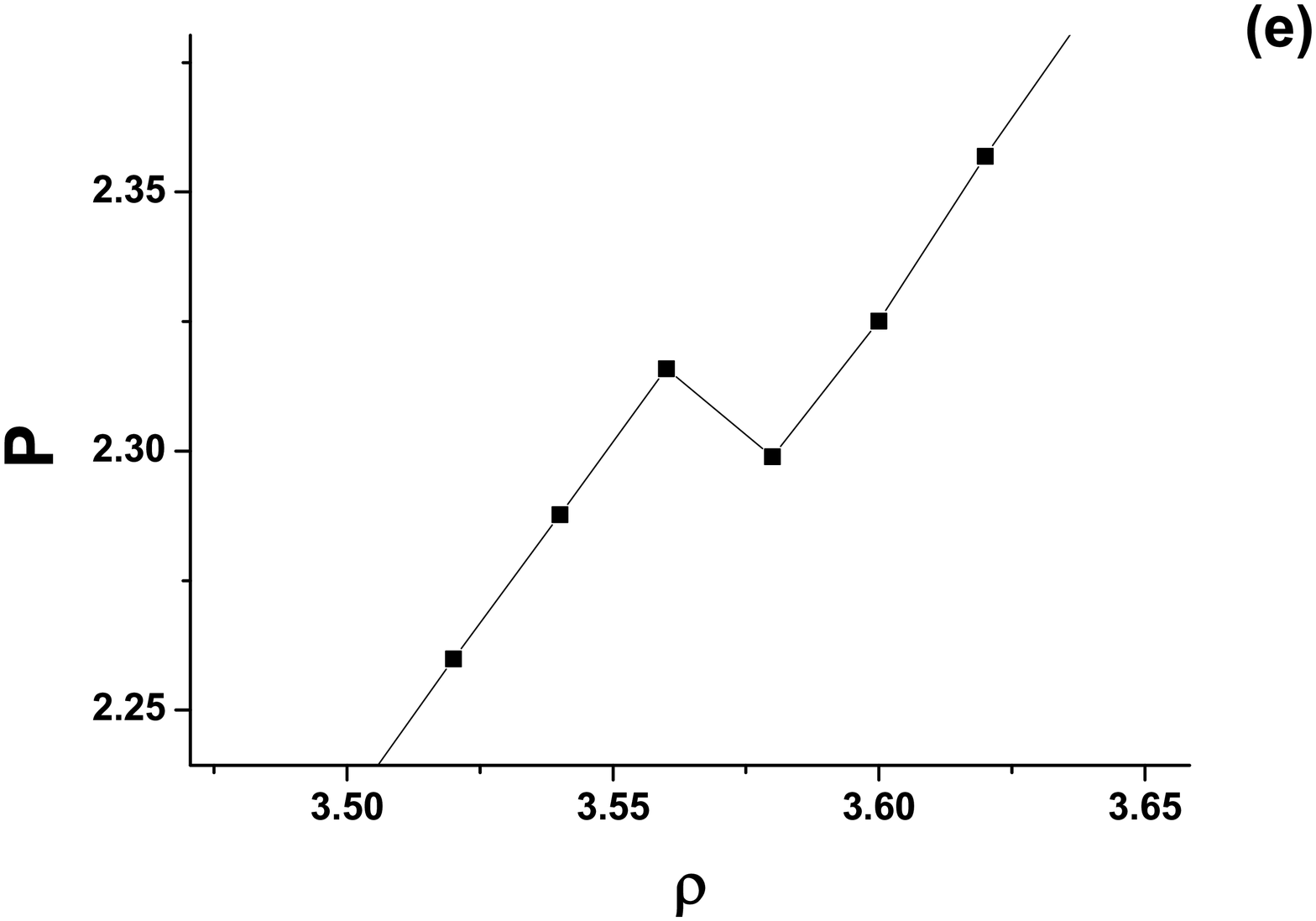}%
\includegraphics[width=4cm,height=4cm]{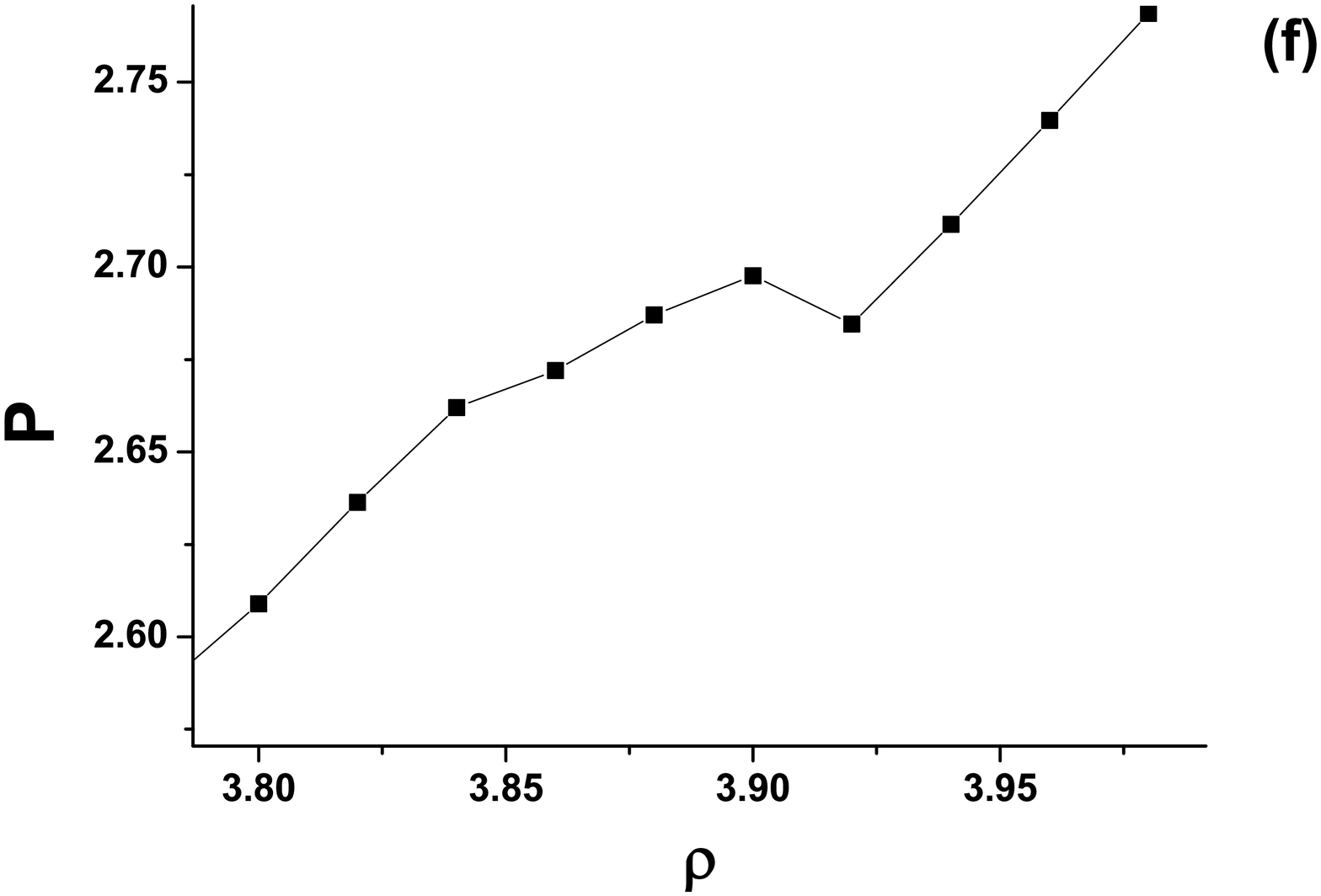}%

\includegraphics[width=4cm,height=4cm]{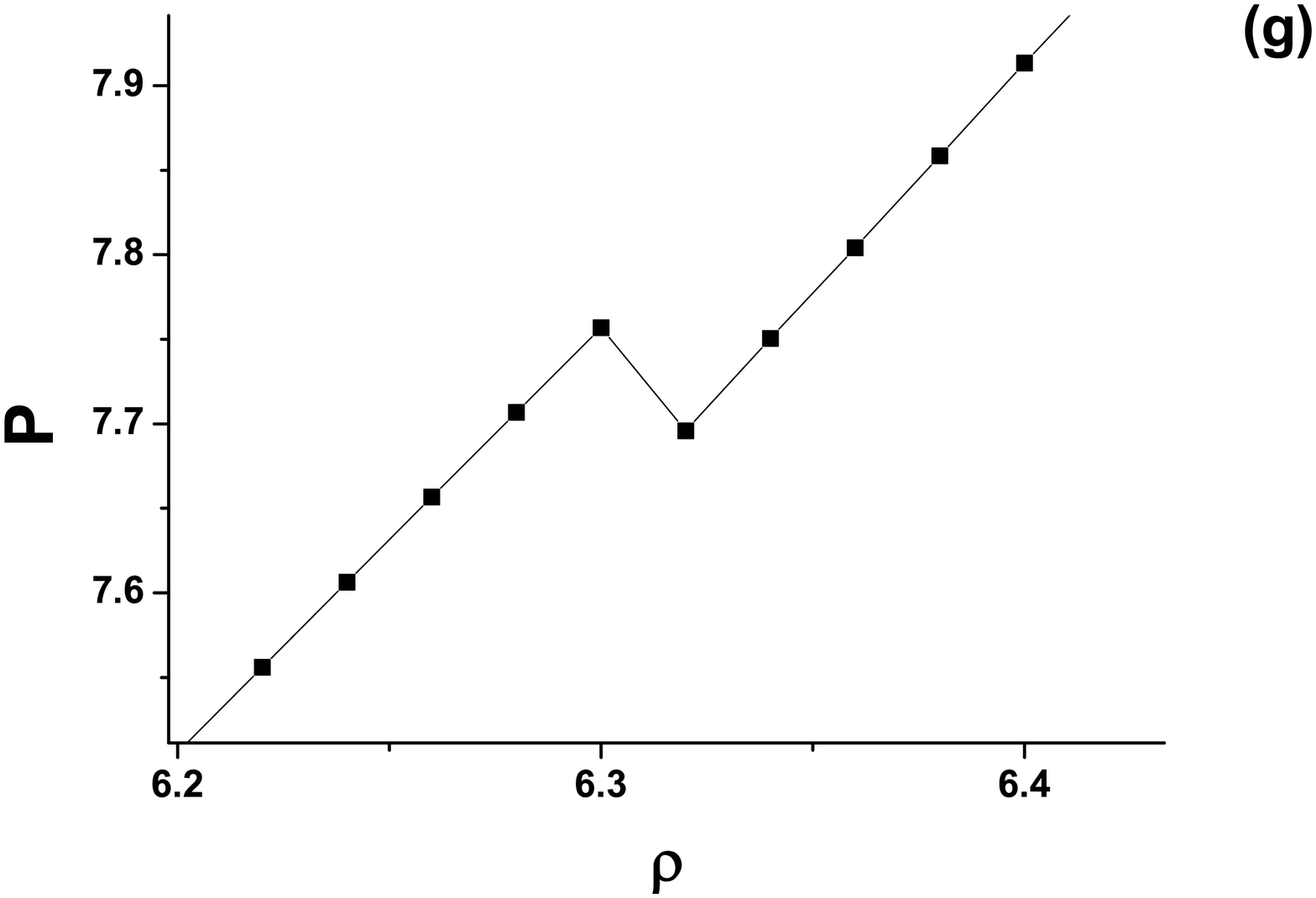}%
\includegraphics[width=4cm,height=4cm]{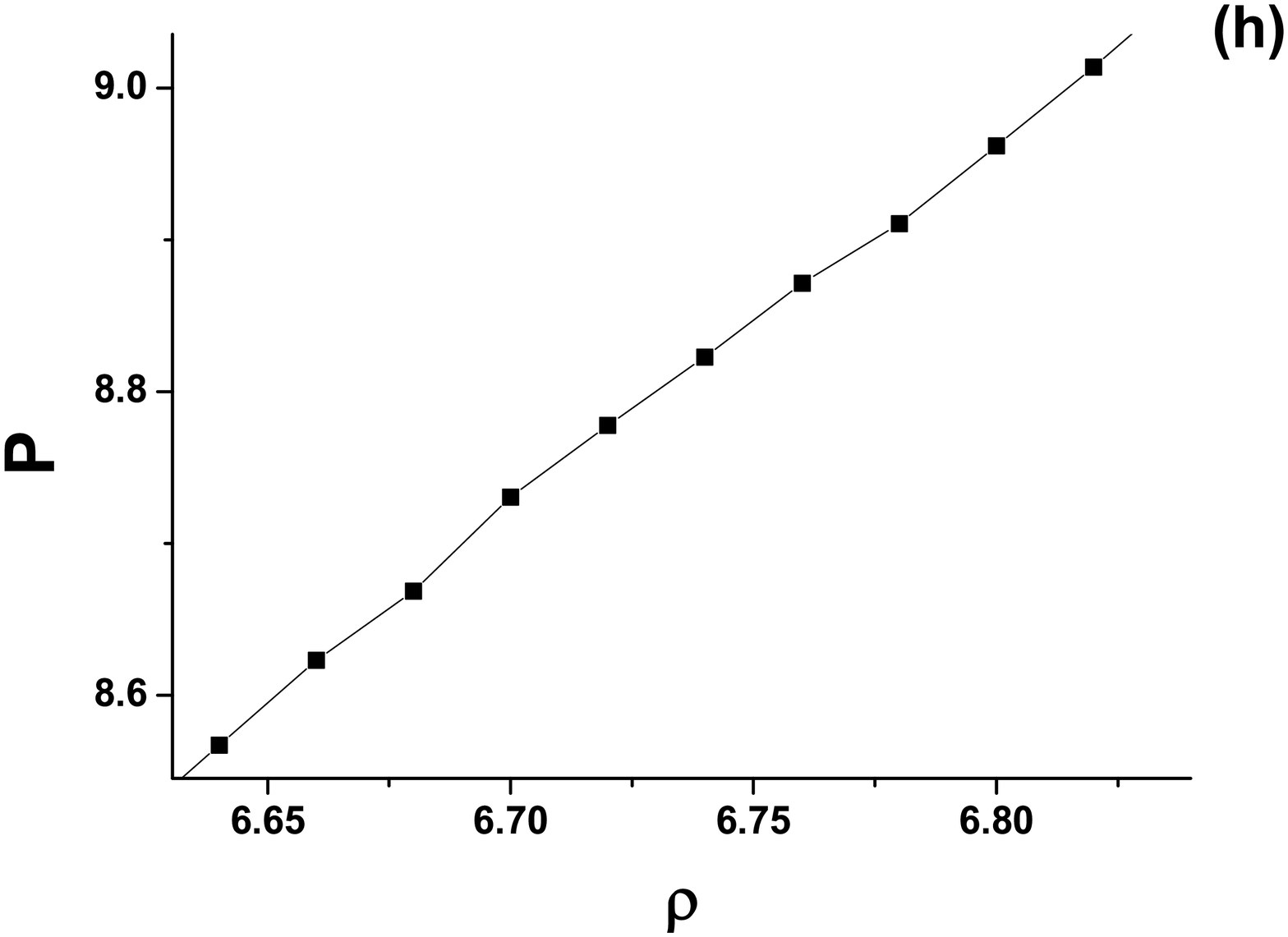}%

\includegraphics[width=4cm,height=4cm]{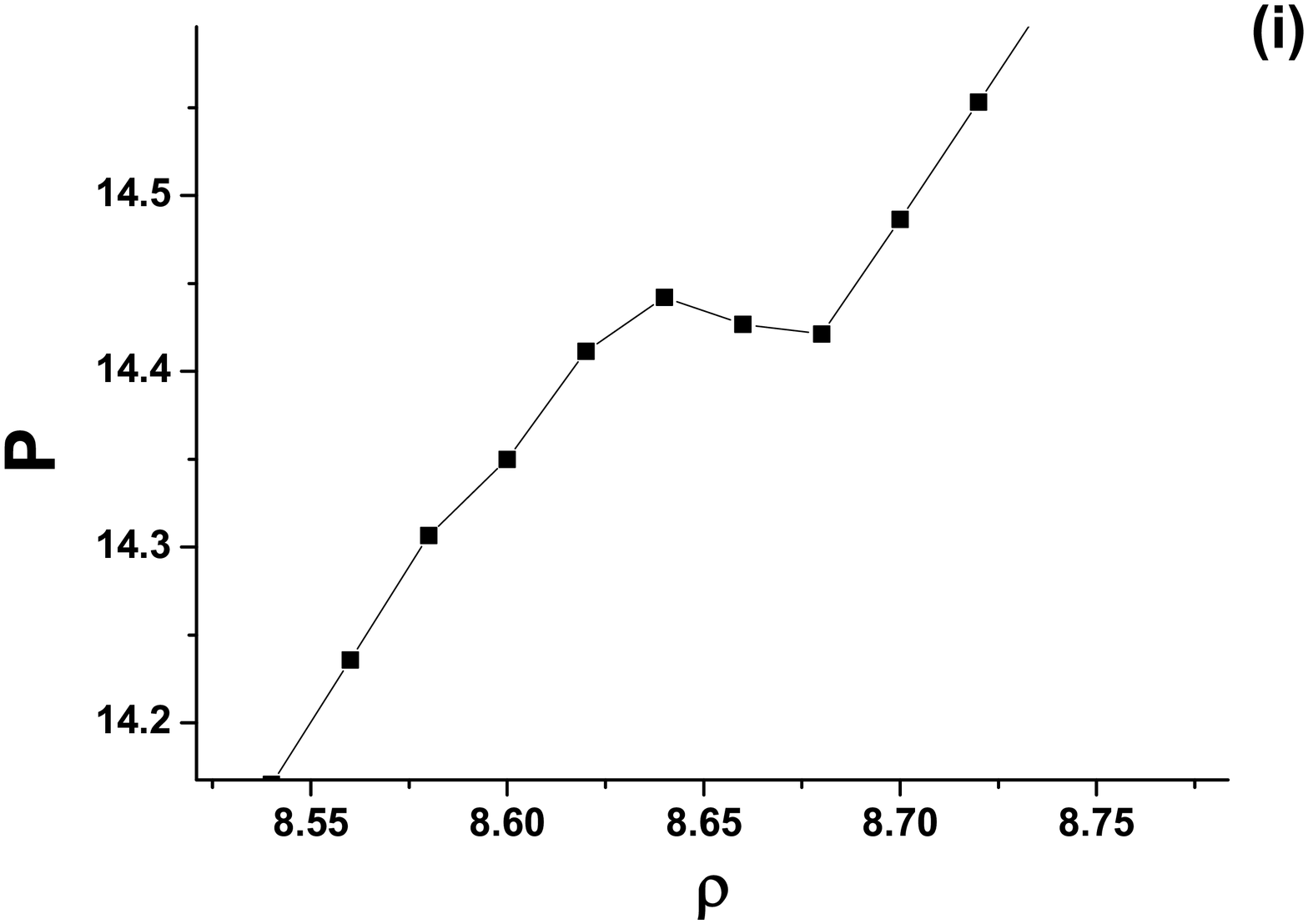}%
\includegraphics[width=4cm,height=4cm]{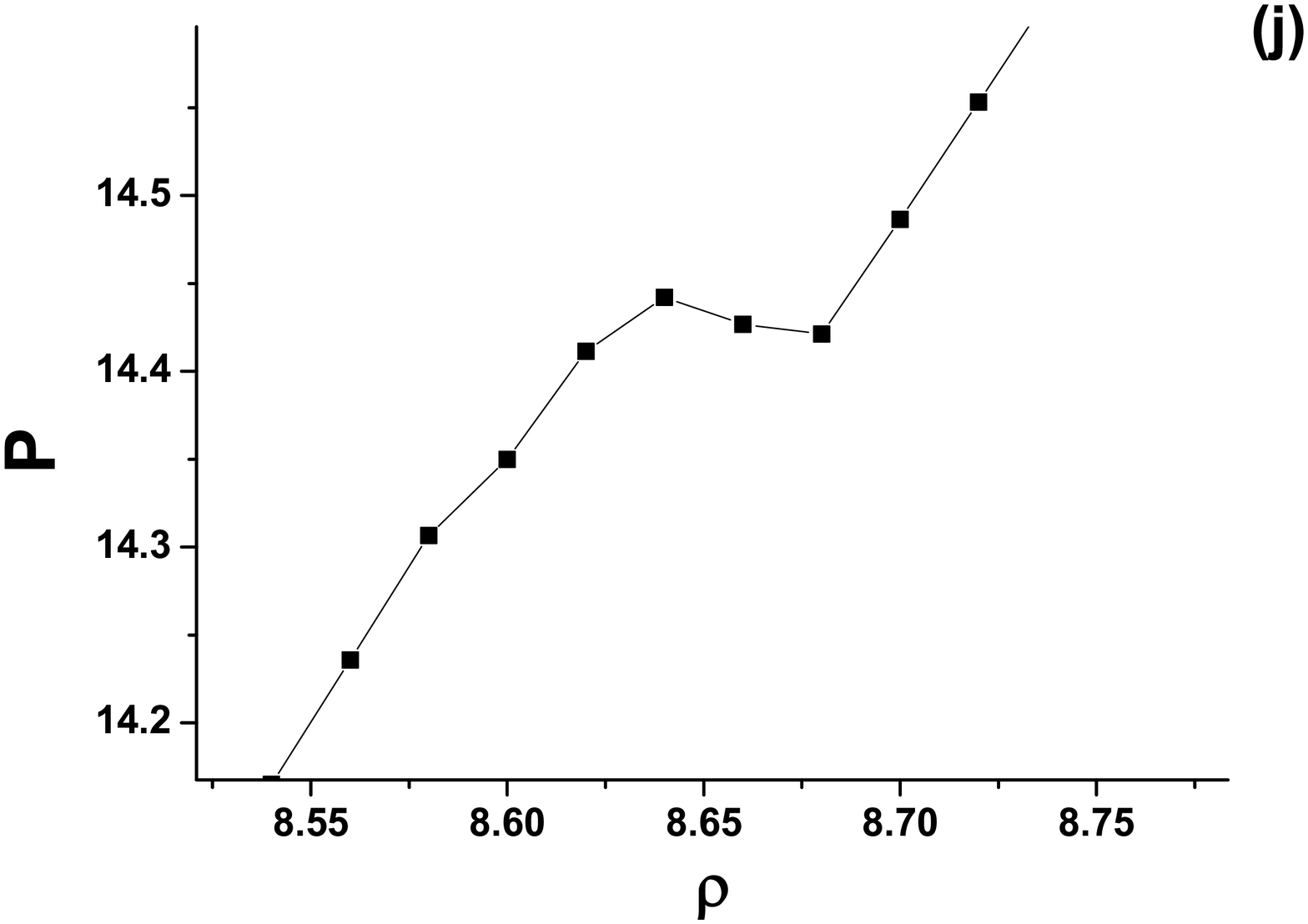}%

\end{center}

\caption{\label{fig:eos} Equation of state of Hertz system at
$T=0.001$. Panel (a) shows the full eos simulated in the present
work. Other panels enlarge the parts of eos in the regions of the
phase transitions.}
\end{figure}

In order to perform a first estimation of the phase diagram of the
system we study its isotherms. Fig. \ref{fig:eos} shows the
isotherm $T=0.001$ (panel (a)). Other panels enlarge the regions
where we observe Mayer-Wood loops which indicate the first order
phase transition. One can see numerous loops which mean that many
phase transitions take place in the system.

\begin{figure}
\begin{center}
\includegraphics[width=10cm,height=9cm]{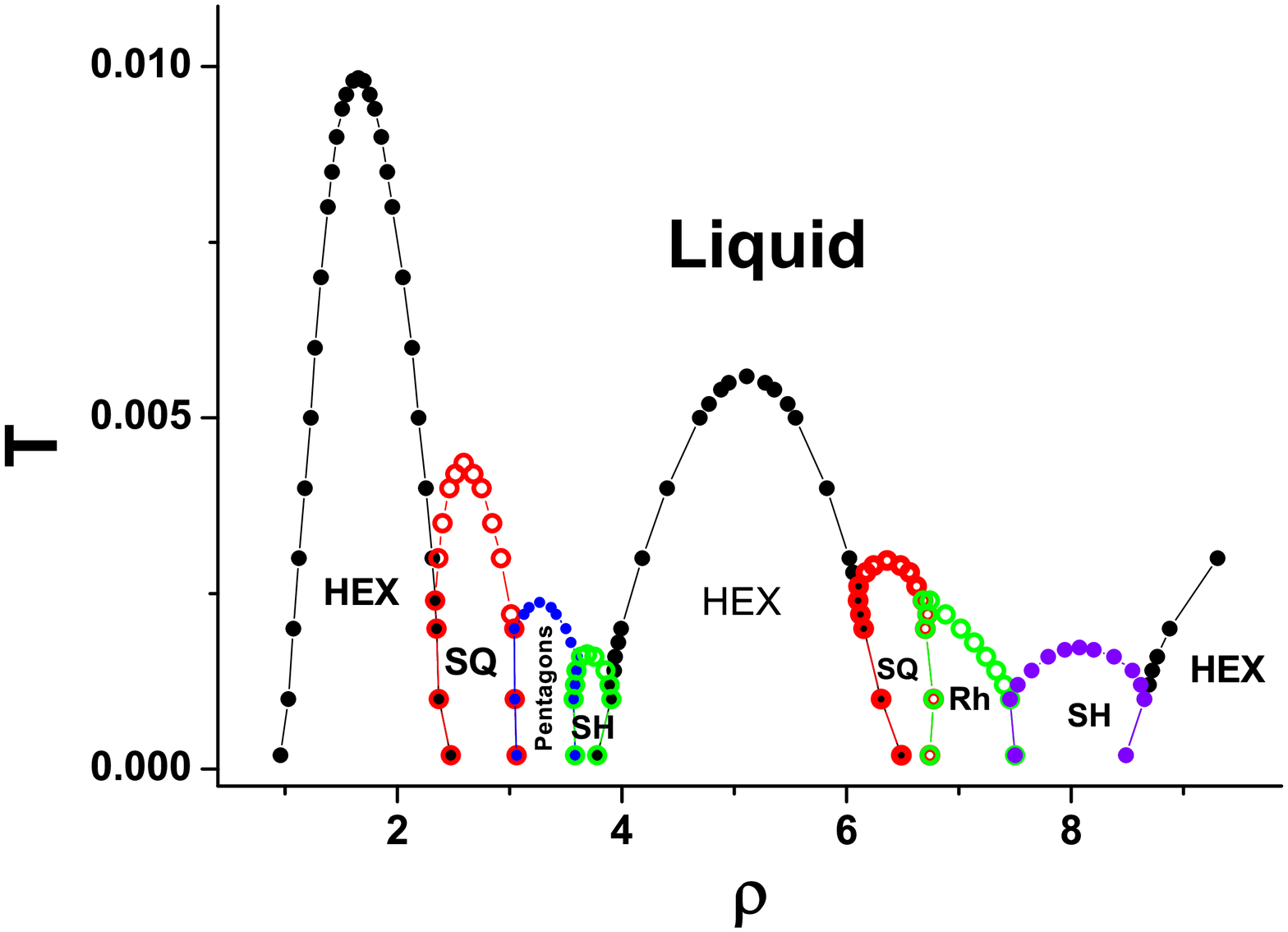}%
\end{center}

\caption{\label{fig:pd-full} Phase diagram of the Hertz system up
to the density $\rho_{max}=10.0$. The notation of the solid phases
is the following: HEX - triangular crystal, SQ - square crystal,
Pentagons - dodecagonal quasicrystal, SH - stretched triangular
phase, RH - rhombohedral crystal. Since the density jumps are
hardly visible in this scale we draw a kind of "average line"
which is in the middle of the transition points. More precise
transition lines for the first two crystal structures (HEX and SQ
with low densities) will be given below. More precise lines of
phase transitions of other phases will be given in subsequent
publications.}
\end{figure}

Using the equations of state along different isotherms we
construct the phase diagram of the system. It is shown on Fig.
\ref{fig:pd-full}. In agreement with the previous studies
\cite{miller,hertzqc} the phase diagram is very complex. We find
that the sequence of phases is different from Ref. \cite{miller}
and corresponds to Ref. \cite{hertzqc}, i.e. in addition to the
triangular and square phases we find also stretched triangular
(SH) and rhombohedral phases. Moreover, since we extend our
investigation to the densities up to $\rho=10.0$ which is above
the upper density in Ref. \cite{hertzqc} we find more phases in
this high density region: the rhombohedral phase transforms into
the second stretched triangular phase and the later transforms
into high density triangular crystal.

We simulated the system at different temperatures in order to find
out the full phase diagram which is shown on Fig.
\ref{fig:pd-full}.

\begin{figure}
\begin{center}
\includegraphics[width=8cm,height=4cm]{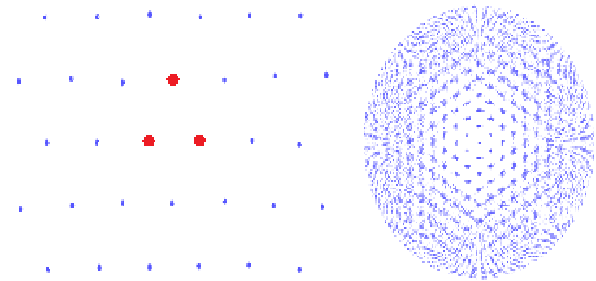}%
\end{center}

\caption{\label{fig:hex-r16} Left panel: example of a snapshot of
the triangular phase with low density. Right panel: diffraction
pattern of this structure. $\rho=1.6$, $T=0.001$.}
\end{figure}

We start from triangular structure with low density. Fig.
\ref{fig:hex-r16} shows a part of configuration of the triangular
phase with low density (left panel) and the corresponding
diffraction pattern.

\begin{figure}
\begin{center}
\includegraphics[width=8cm,height=4cm]{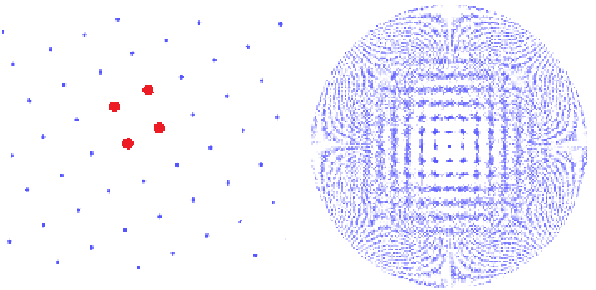}%
\end{center}

\caption{\label{fig:sq-r28} Left panel: example of a snapshot of
the square crystal with low density. Right panel: diffraction
pattern of this structure. $\rho=2.8$, $T=0.001$.}
\end{figure}

Analogously Fig. \ref{fig:sq-r28} shows an example of snapshot and
diffraction pattern of square phase with low density.

\begin{figure}
\begin{center}
\includegraphics[width=8cm,height=4cm]{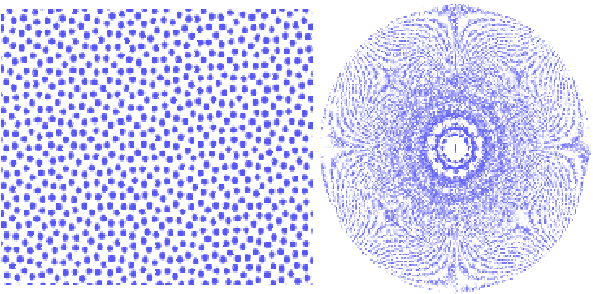}%
\end{center}

\caption{\label{fig:pent-34} Left panel: example of a snapshot of
the dodecagonal quasicrystal with low density. Right panel:
diffraction pattern of this structure. $\rho=3.4$, $T=0.001$.}
\end{figure}

While the triangular and square phases are relatively easy to
identify, the next phase is the one consisting of pentagons. In
Ref. \cite{miller} this structure was not identified. The authors
proposed a very complex way of description of this structure by
decomposing it into five sublattices. The authors of Ref.
\cite{hertzqc} defined this structure as dodecagonal quasicrystal.
Our study confirms this finding. It can be realized from the
diffraction pattern shown in Fig. \ref{fig:pent-34} which clearly
demonstrates 12-fold symmetry corresponding to the dodecagonal
quasicrystal.

\begin{figure}
\begin{center}
\includegraphics[width=8cm,height=4cm]{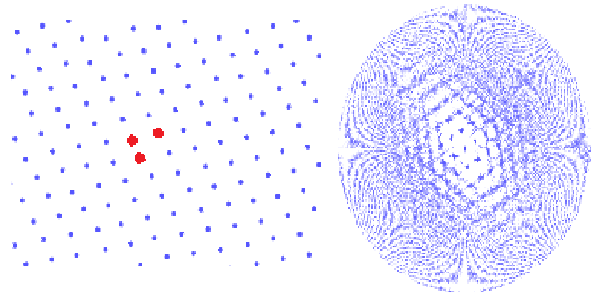}%
\end{center}

\caption{\label{fig:sh-37} Left panel: example of a snapshot of
the stretched triangular crystal phase with low density. Right
panel: diffraction pattern of this structure. $\rho=3.7$,
$T=0.001$.}
\end{figure}

The next structure we identify as stretched triangular structure.
This can be visible from the snapshot (Fig. \ref{fig:sh-37}). But
most clearly it can be seen from the diffraction pattern given on
the right panel of the same figure.

\begin{figure}
\begin{center}
\includegraphics[width=8cm,height=4cm]{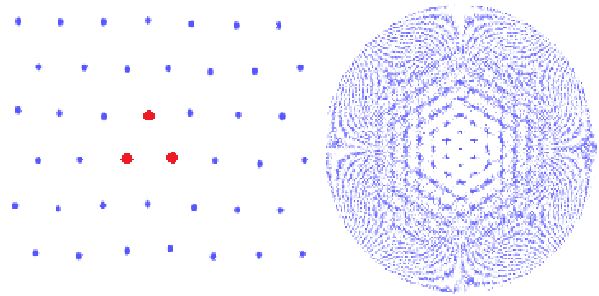}%
\end{center}

\caption{\label{fig:hex-52} Left panel: example of a snapshot of
the triangular phase with intermediate density. Right panel:
diffraction pattern of this structure. $\rho=5.2$, $T=0.001$.}
\end{figure}

\begin{figure}
\begin{center}
\includegraphics[width=8cm,height=4cm]{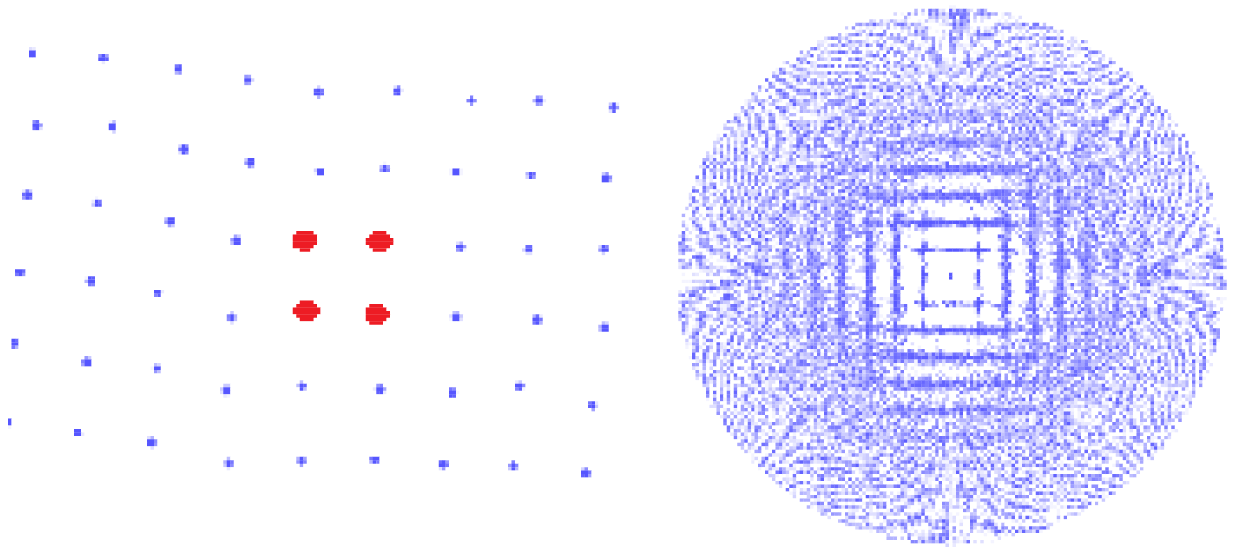}%
\end{center}

\caption{\label{fig:sq-66} Left panel: example of a snapshot of
the square phase with intermediate density. Right panel:
diffraction pattern of this structure. $\rho=6.6$, $T=0.001$.}
\end{figure}

\begin{figure}
\begin{center}
\includegraphics[width=8cm,height=4cm]{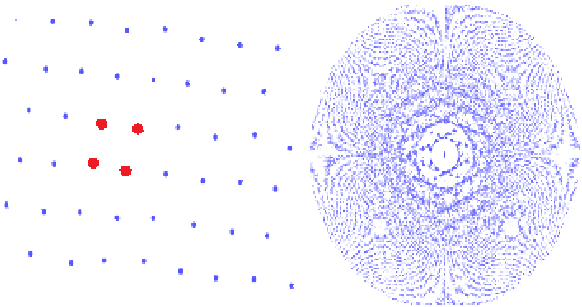}%
\end{center}

\caption{\label{fig:rh-72} Left panel: example of a snapshot of
the rhombohedral phase. Right panel: diffraction pattern of this
structure. $\rho=7.2$, $T=0.001$.}
\end{figure}

\begin{figure}
\begin{center}
\includegraphics[width=8cm,height=4cm]{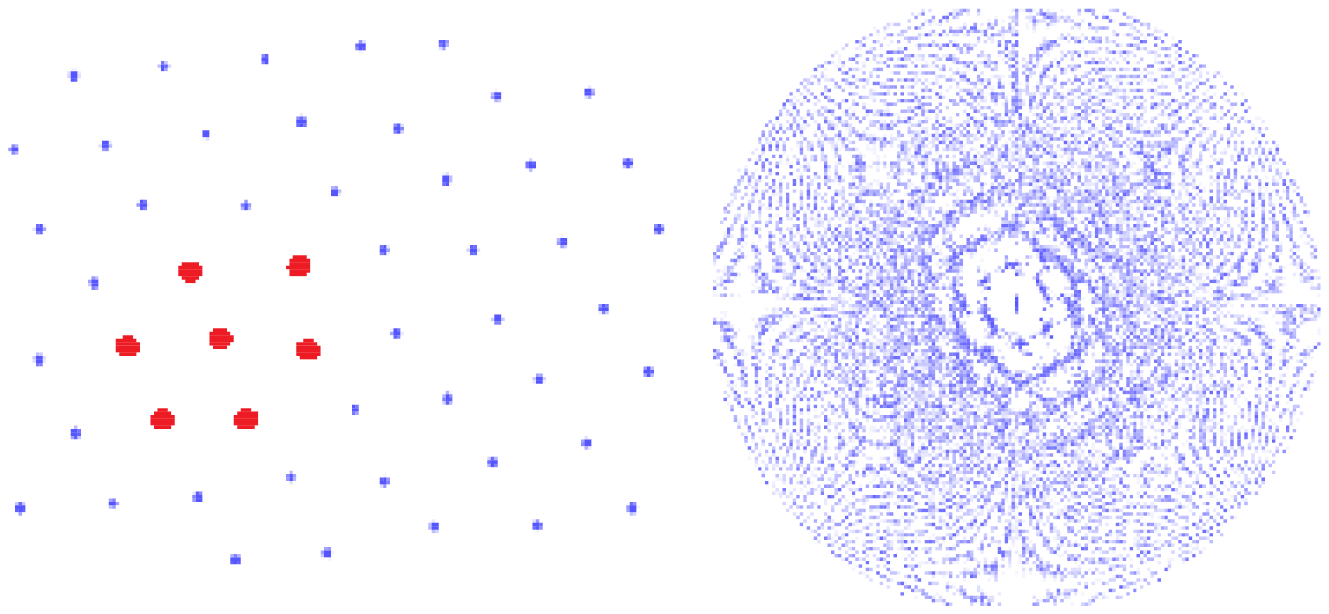}%
\end{center}

\caption{\label{fig:sh-8} Left panel: example of a snapshot of the
stretched triangular phase with high density. Right panel:
diffraction pattern of this structure. $\rho=8.0$, $T=0.001$.}
\end{figure}

\begin{figure}
\begin{center}
\includegraphics[width=8cm,height=4cm]{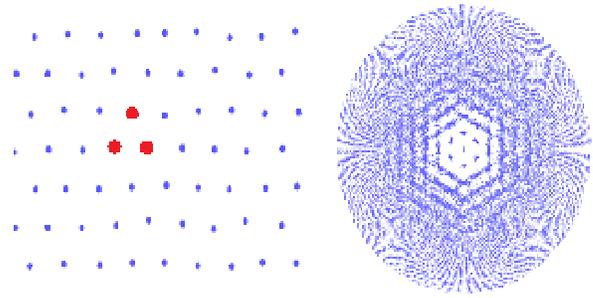}%
\end{center}

\caption{\label{fig:hex-9} Left panel: example of a snapshot of
the triangular phase with high density. Right panel: diffraction
pattern of this structure. $\rho=9.0$, $T=0.001$.}
\end{figure}

Further increase of the density leads to almost exact repetition
of the sequence of phases, but the quasicrystalline phase changes
to the rhombohedral crystal. The sequence of the phases becomes
the following: triangular (Fig. \ref{fig:hex-52}), square (Fig.
\ref{fig:sq-66}), rhombohedral (Fig. \ref{fig:rh-72}), stretched
triangular (Fig. \ref{fig:sh-8}) and finally again triangular
(Fig. \ref{fig:hex-9}). It is worth to note that the stretched
triangular and rhombohedral structures do not look very obvious
from snapshots and radial distribution functions, but the
structure becomes apparent from the diffraction patters.

\section{Low density part of the phase diagram}

Now we consider in more details the low density part of the phase
diagram, which includes the first two crystalline phases
(triangular and square) and liquid.

\begin{figure}
\begin{center}
\includegraphics[width=6cm,height=6cm]{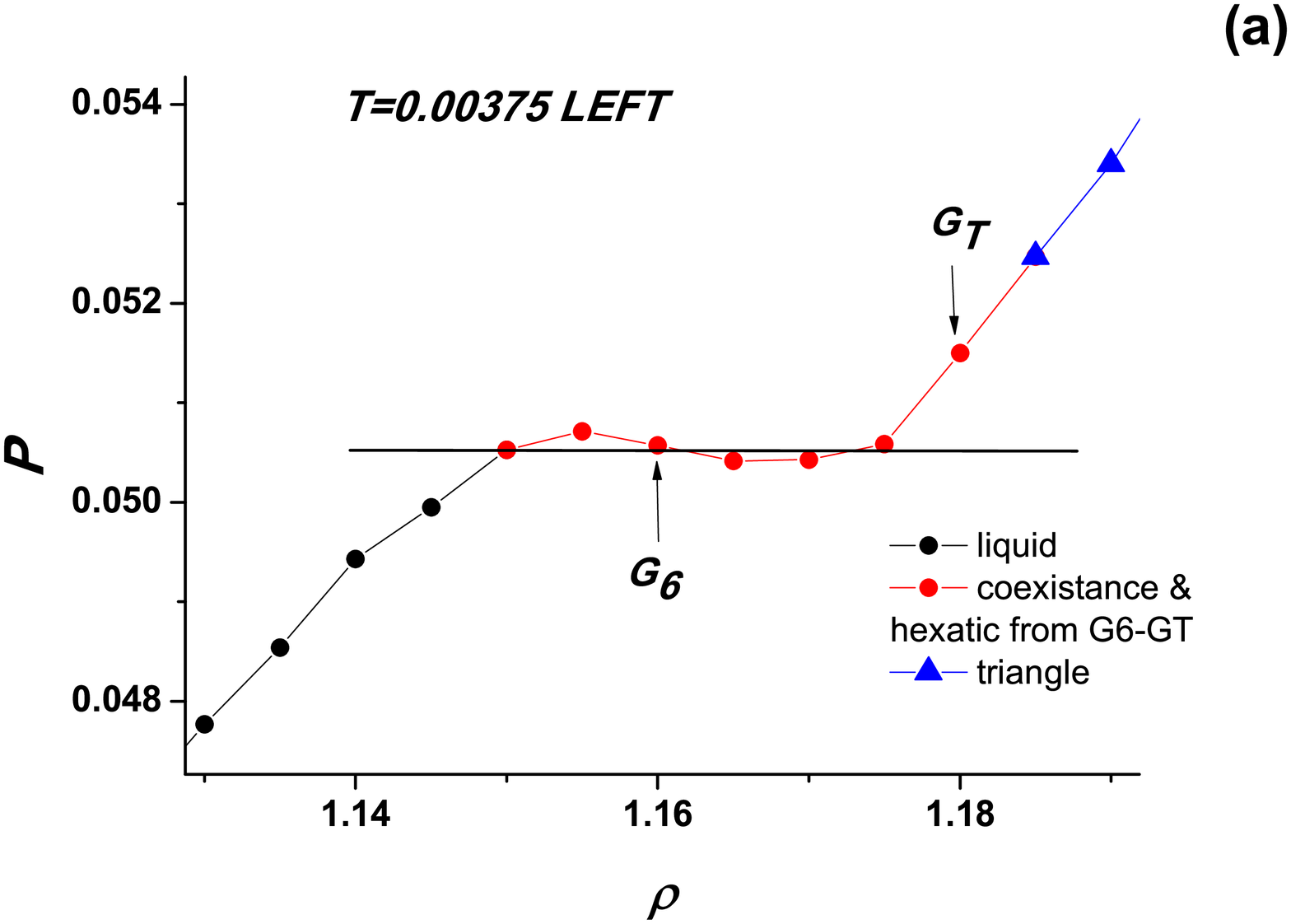}%

\includegraphics[width=6cm,height=6cm]{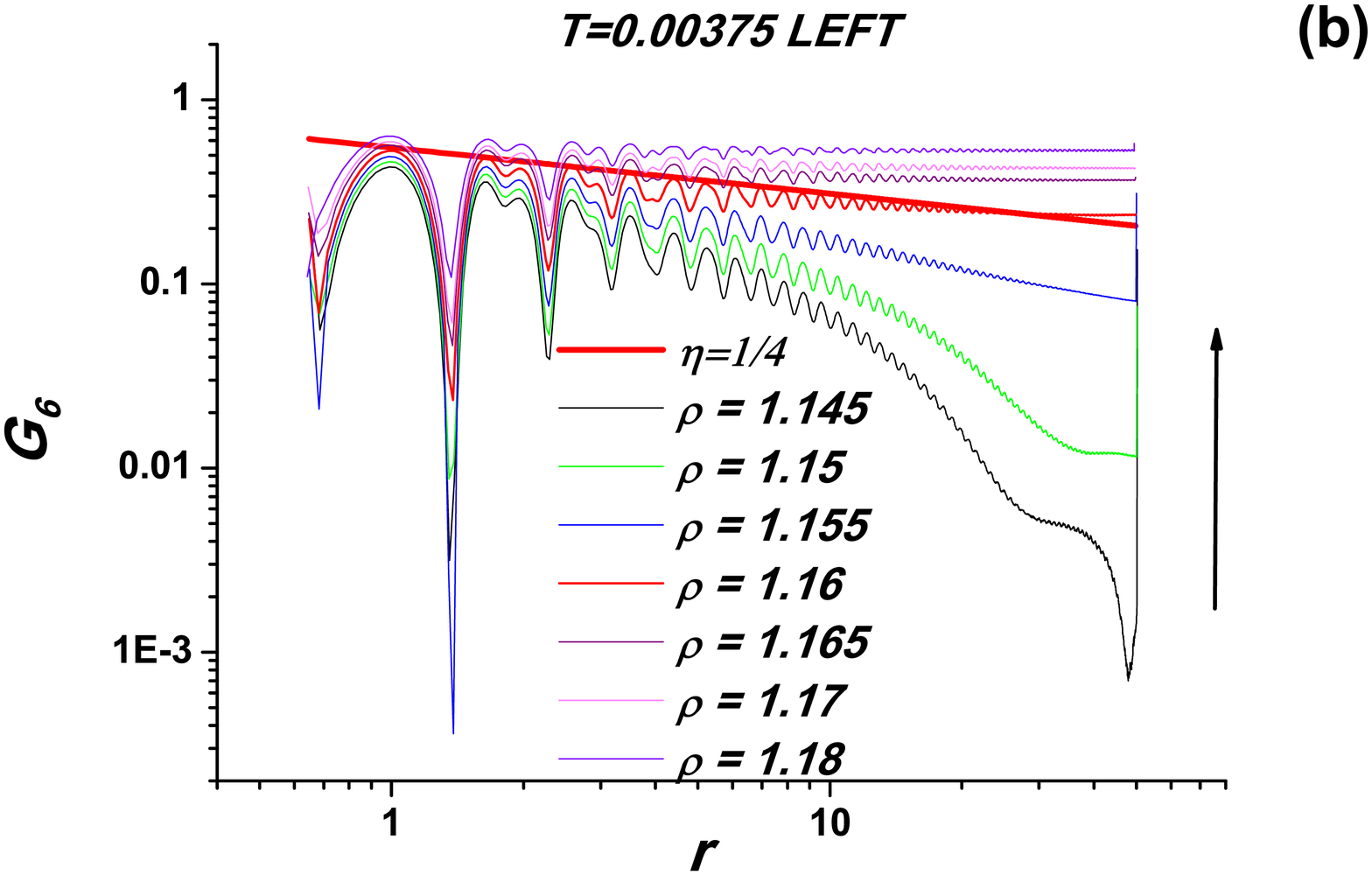}%

\includegraphics[width=6cm,height=6cm]{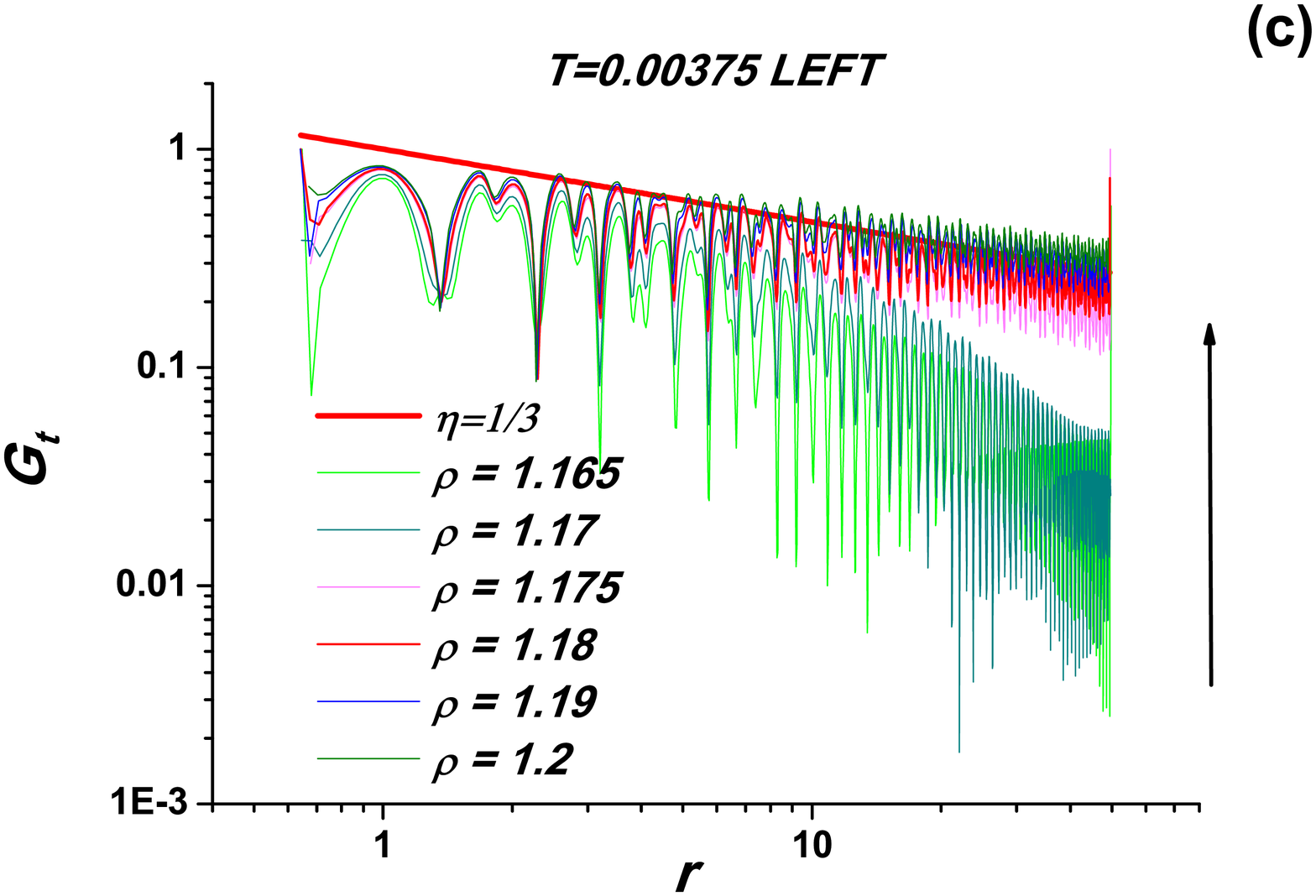}%

\end{center}

\caption{\label{fig:hex-left-eos} (a) Equation of state of
Hertzian spheres at $T=0.0375$ in the region of liquid to low
density triangular crystal transition at lower densities (left
branch of the melting line in density-temperature plane). Point
$G_T$ corresponds to the density at which transition from crystal
to hexatic phase takes place according to the criterion
$\eta_T=1/3$. Point $G_6$ marks the transition from hexatic phase
to isotropic liquid from the criterion $\eta_6=1/4$. (b) The
behavior of orientational correlation function $G_6$ of the same
system. (c) The behavior of translational correlation function
$G_T$ of the same system. The arrows in the panels (b) and (c)
show the direction of the density increase.}
\end{figure}

Fig. \ref{fig:hex-left-eos} (a) shows an example of equation of
state at the low density branch of the melting line of the
low-density triangular phase. One can see that the equation of
state demonstrates a Mayer-Wood loop which means that a first
order transition takes place. By analyzing the orientational and
translational correlation functions shown in Figs.
\ref{fig:hex-left-eos} (b) and (c), we conclude that the hexatic
phase transforms into the isotropic liquid through first order
transition. As higher densities there is a transition from hexatic
phase into triangular solid. This kind of melting corresponds to
the scenario proposed in Refs.
\cite{foh1,foh2,foh3,foh4,foh5,foh6} and is in agreement with the
results of Ref. \cite{hertzmelt}.

The next Fig. \ref{fig:hex-right-eos} (a)-(c) demonstrates the
same analysis for the right branch of the melting line. No
Mayer-Wood loop is observed here at high temperatures. By
analyzing the equation of state and the correlation functions we
conclude that the system melts in accordance to the BKTHNY
scenario which is again consistent with Ref. \cite{hertzmelt}.

\begin{figure}
\begin{center}
\includegraphics[width=6cm,height=6cm]{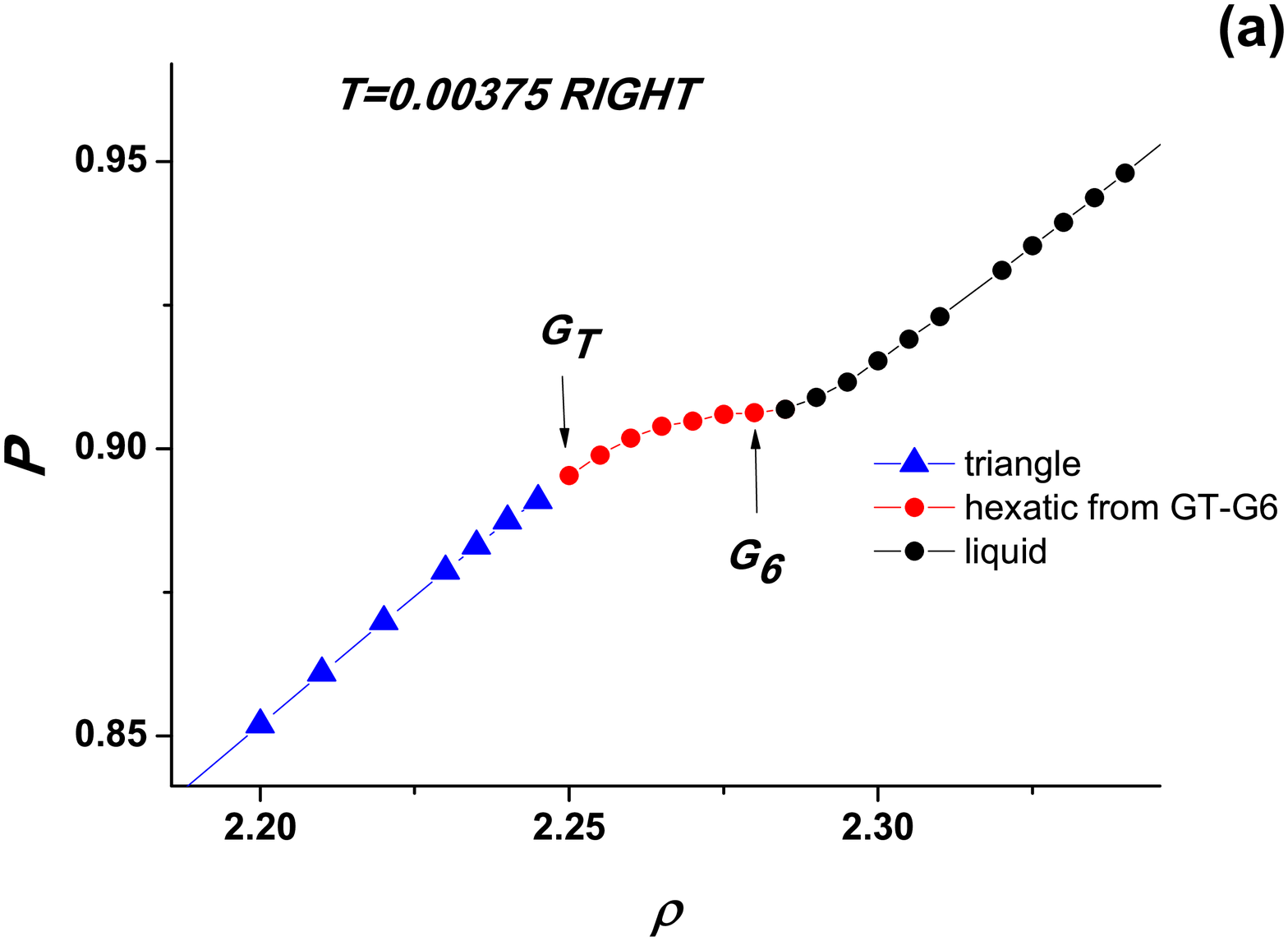}%

\includegraphics[width=6cm,height=6cm]{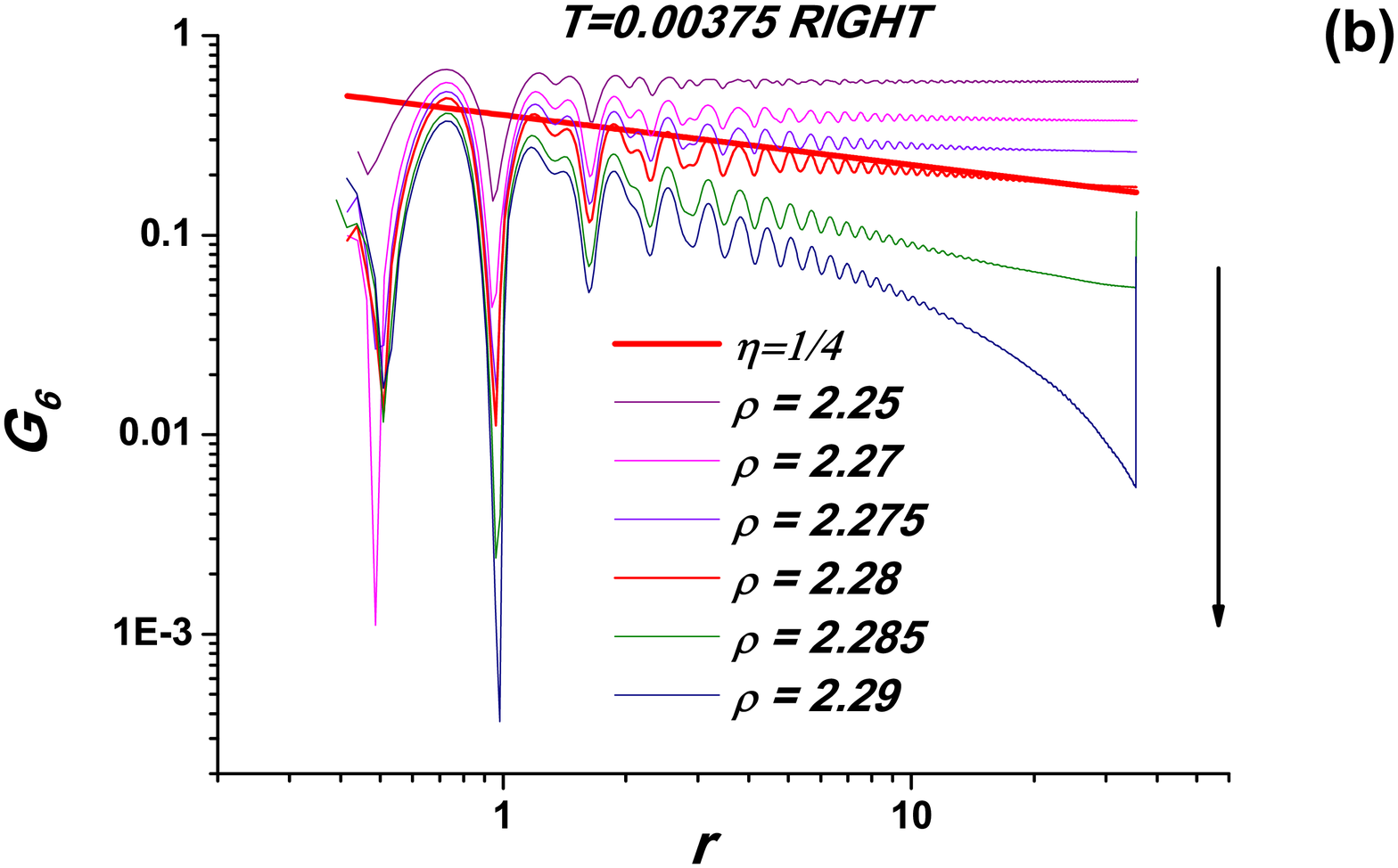}%

\includegraphics[width=6cm,height=6cm]{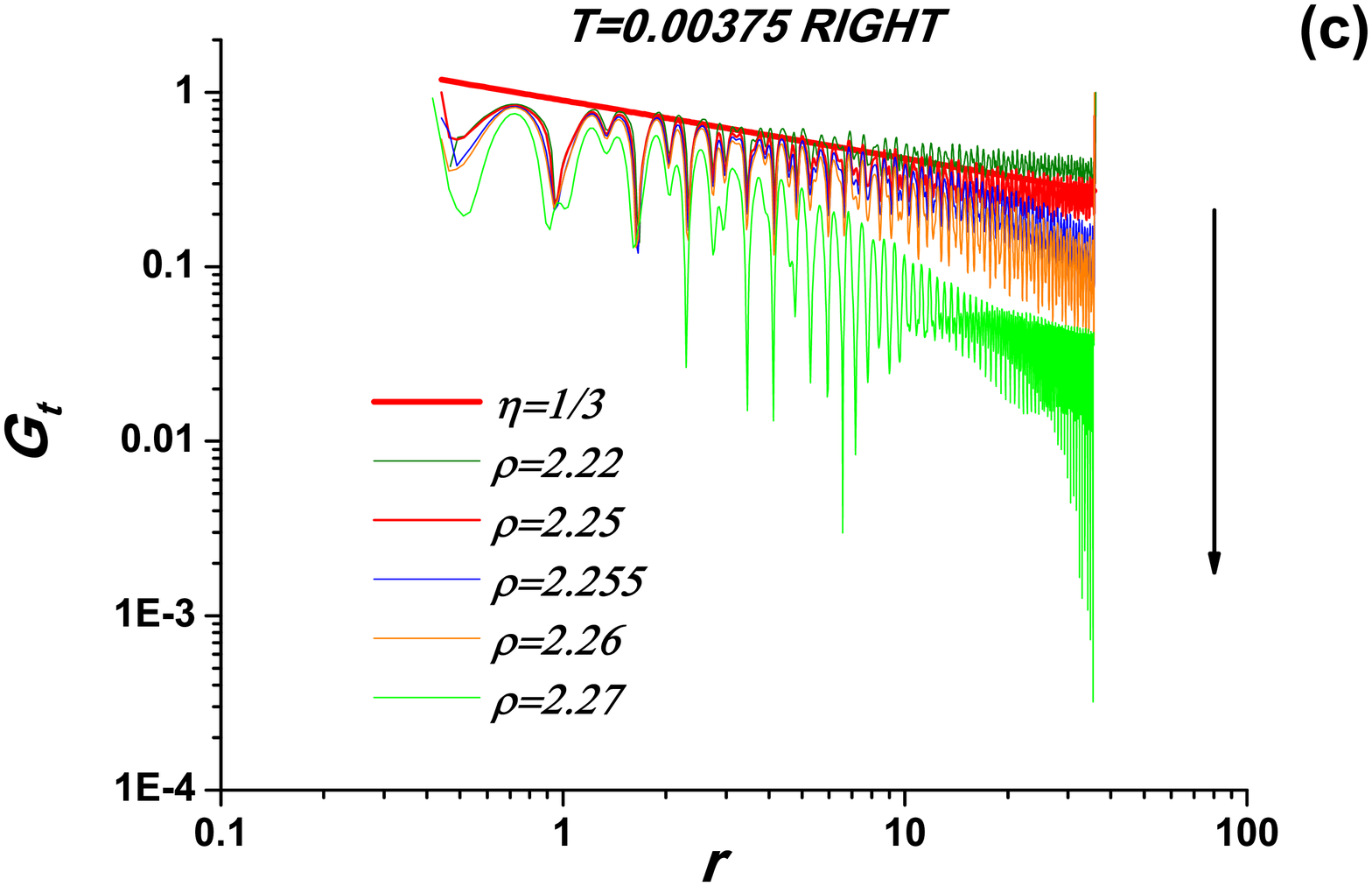}%

\end{center}

\caption{\label{fig:hex-right-eos} (a) Equation of state of
Hertzian spheres at $T=0.0375$ in the region of liquid to low
density triangular crystal transition (right branch of the melting
line in density-temperature plane). Point $G_T$ corresponds to the
density at which transition from crystal to hexatic phase takes
place according to the criterion $\eta_T=1/3$. Point $G_6$ marks
the transition from hexatic phase to isotropic liquid from the
criterion $\eta_6=1/4$. (b) The behavior of orientational
correlation function $G_6$ of the same system. (c) The behavior of
translational correlation function $G_T$ of the same system. The
arrows in the panels (b) and (c) show the direction of the density
increase.}
\end{figure}

\begin{figure}
\begin{center}
\includegraphics[width=6cm,height=6cm]{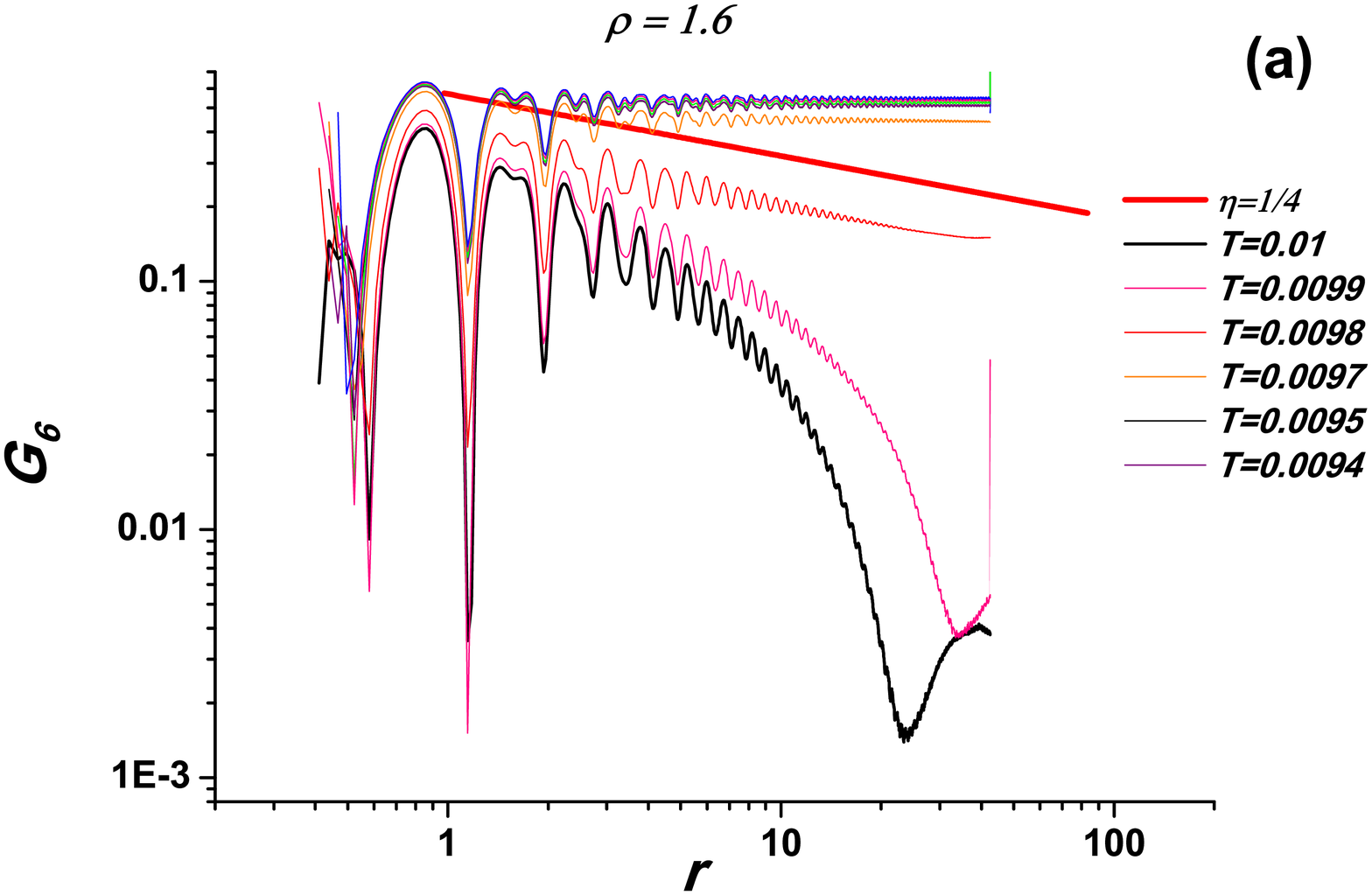}%

\includegraphics[width=6cm,height=6cm]{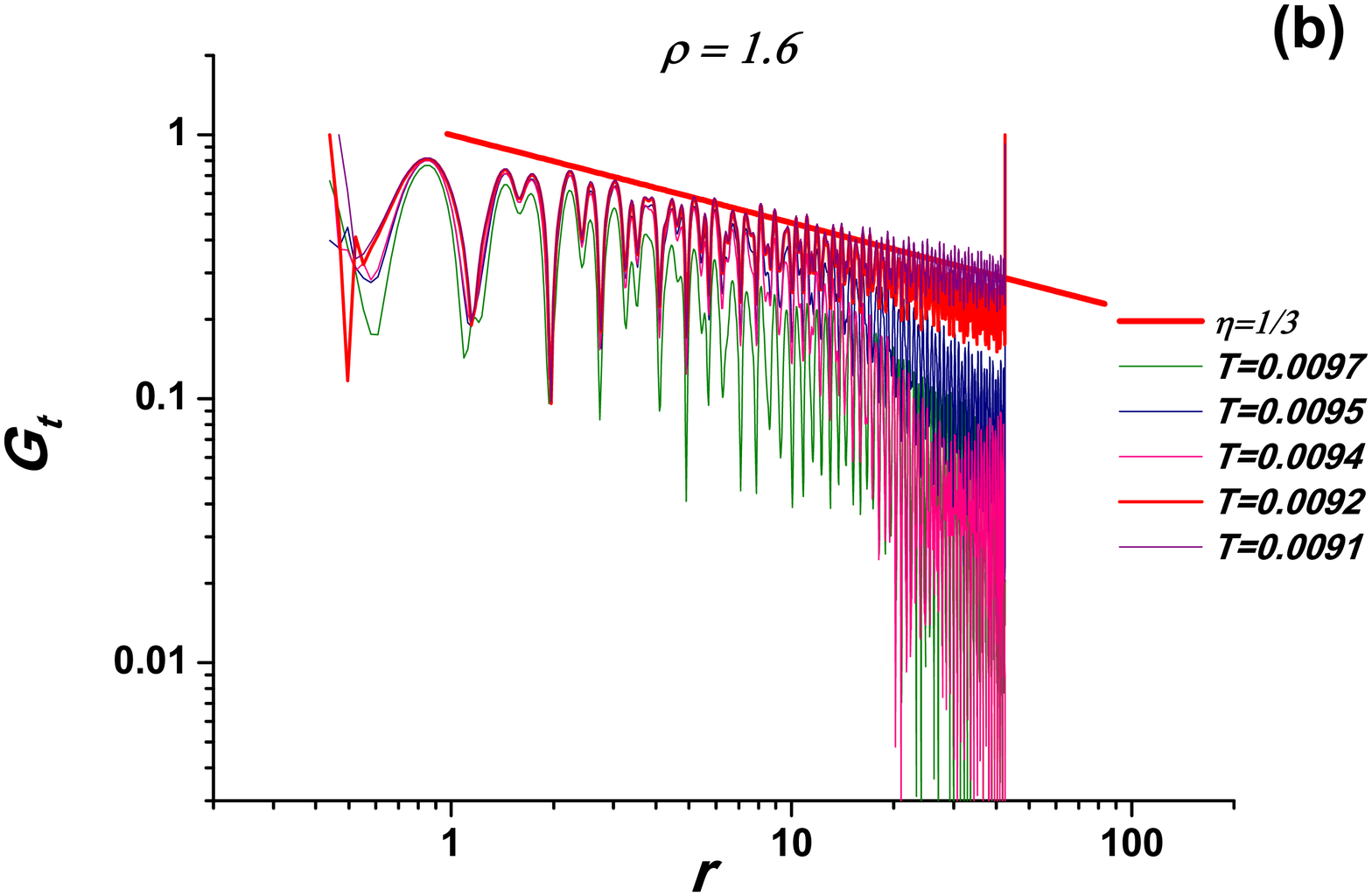}%

\end{center}

\caption{\label{fig:g6gtr16} (a) The behavior of orientational
correlation function $G_6$ at $\rho=1.6$ and different
temperatures. (c) The behavior of translational correlation
function $G_T$ of the same system. }
\end{figure}

Unlike the Ref. \cite{hertzmelt} we find that at the maximum on
the melting line the melting still goes through the hexatic phase.
Fig. \ref{fig:g6gtr16} shows the behavior of the orientational and
translational correlation functions at the maximum of the melting
line. From these plots one can see that from the condition
$\eta_T=1/3$ the transition from crystal into hexatic phase takes
place at $T=0.0092$ while from the condition $\eta_6=1/4$ the
transition of the hexatic phase into isotropic liquid appears at
$T=0.0098$.

\begin{figure}
\begin{center}
\includegraphics[width=6cm,height=6cm]{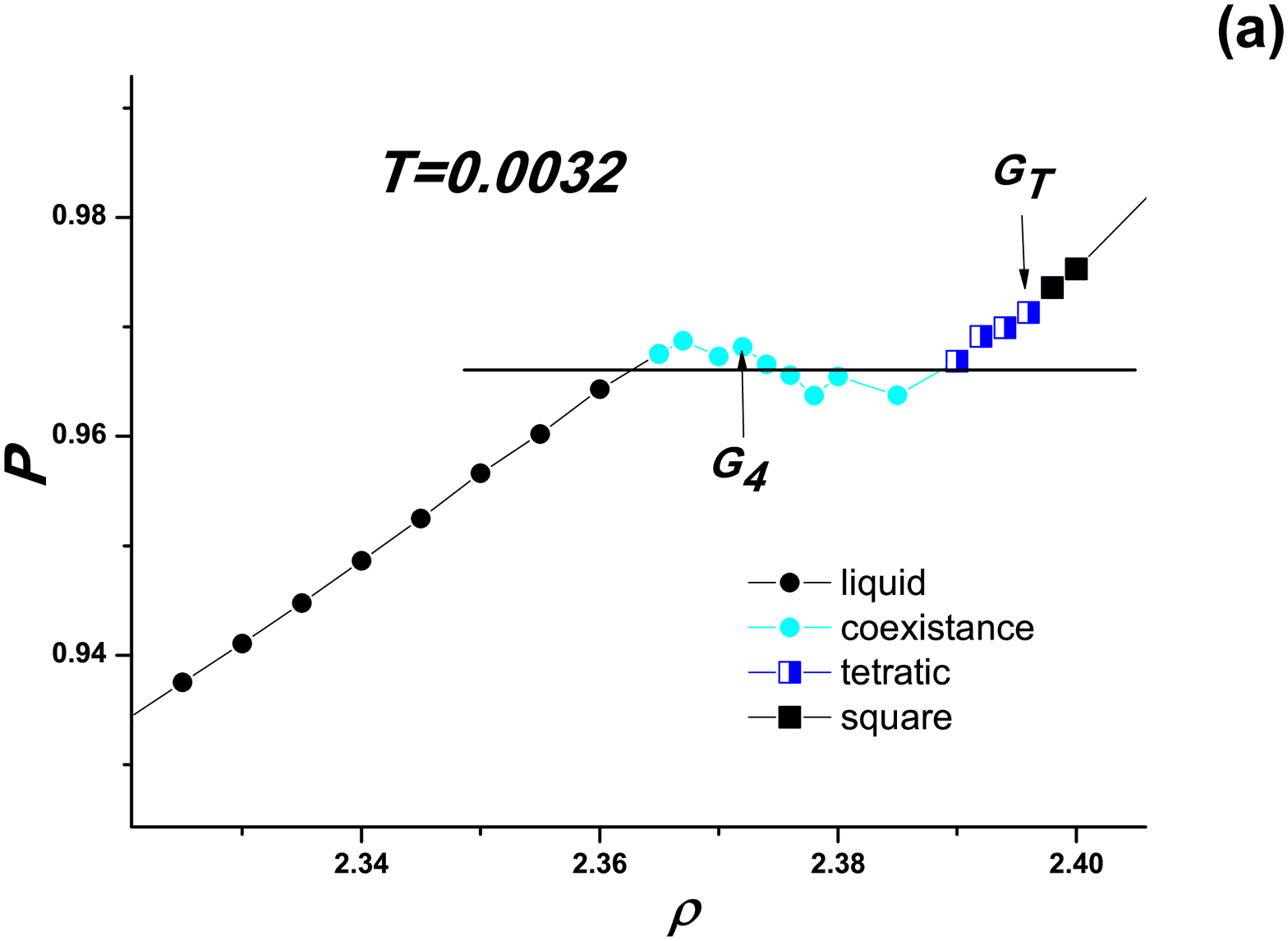}%

\includegraphics[width=6cm,height=6cm]{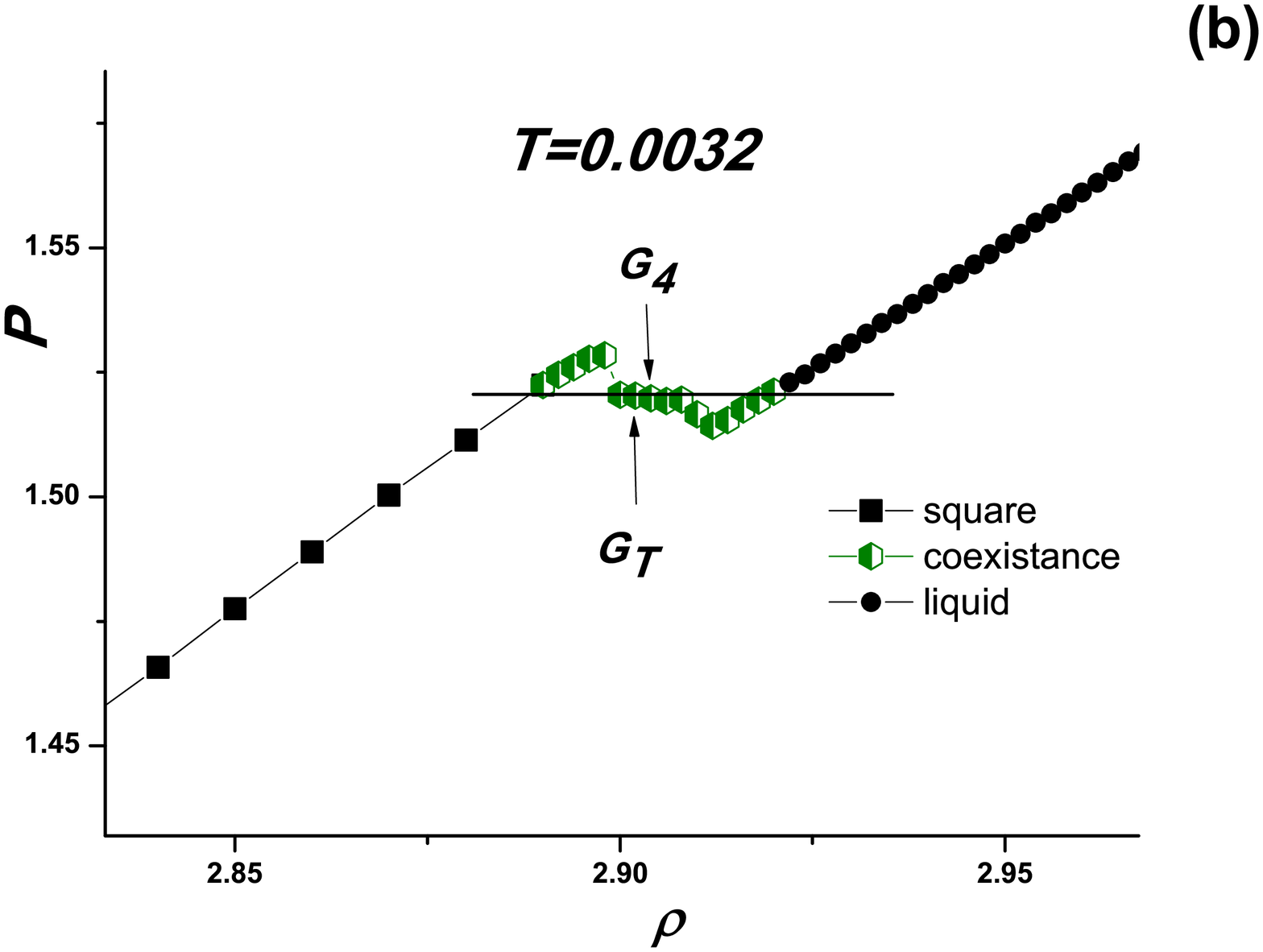}%

\end{center}

\caption{\label{fig:eos-sq} Equation of state of Hertzian spheres
at $T=0.0032$ in the region crossing the borders of the square
crystal phase. Panel  (a) represents the left branch while panel
(b) the right one. The arrows mark the transitions from crystal to
tetractic phase ($G_T$) and from tetratic phase into isotropic
liquid.}
\end{figure}

Similar analysis was performed also for the square crystal. Fig.
\ref{fig:eos-sq} shows the equation of state of Hertzian disks at
$T=0.0032$ and the densities interval crossing the square crystal.
The qualitative behavior of the square crystal looks the same as
the triangular one. At the left branch there is a Mayer-Wood loop
between the isotropic liquid and tetratic phase and the tetratic
phase itself transforms into crystal via a continuous transition.
At the right branch we observe the Mayer-Wood loop and both $G_6$
and $G_T$ criteria fall into the region of the loop. It states
that the system melts through the first order phase transition at
the right branch of the melting curve of the square crystal.

\begin{figure}
\begin{center}
\includegraphics[width=6cm,height=6cm]{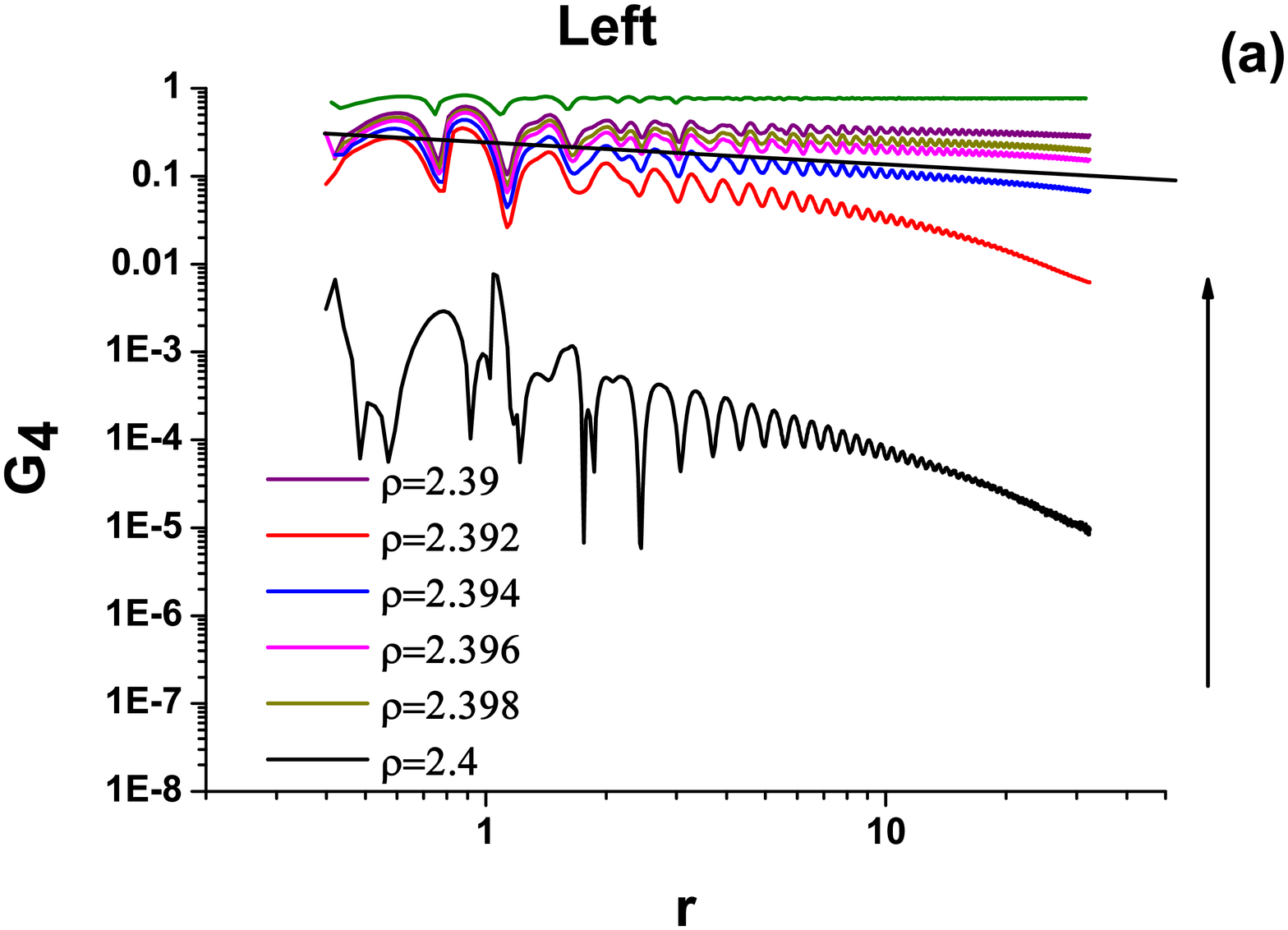}%

\includegraphics[width=6cm,height=6cm]{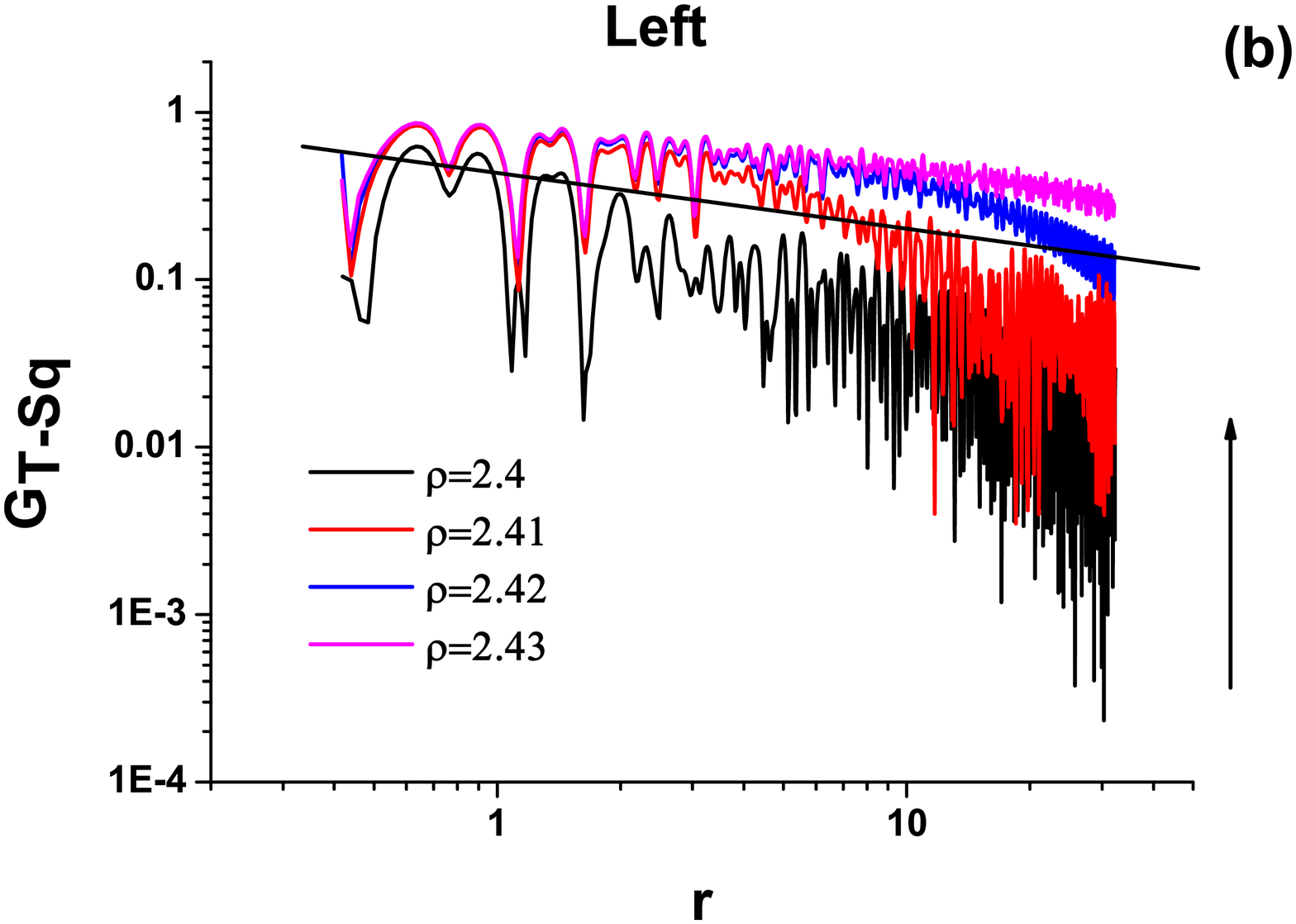}%

\includegraphics[width=6cm,height=6cm]{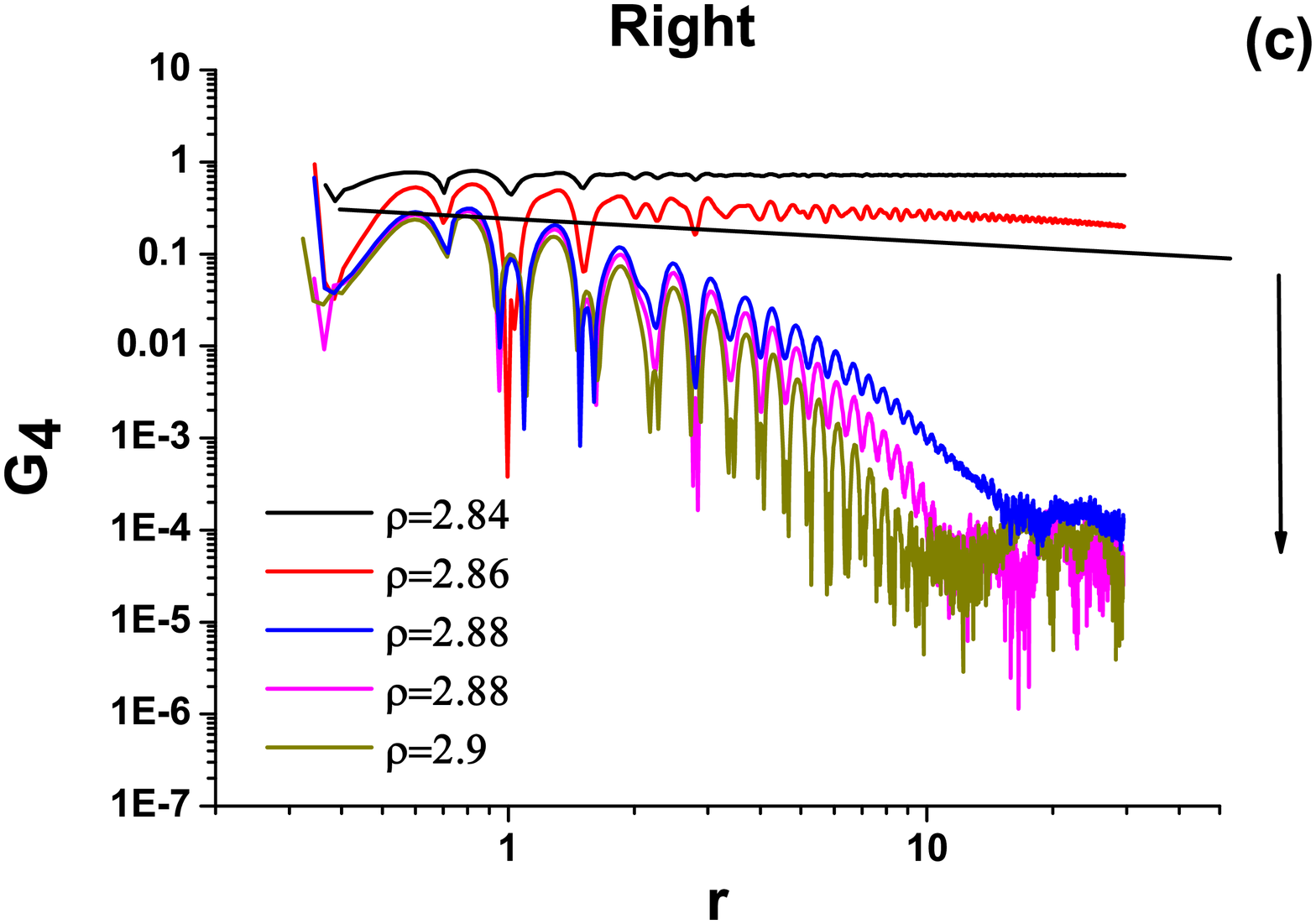}%

\includegraphics[width=6cm,height=6cm]{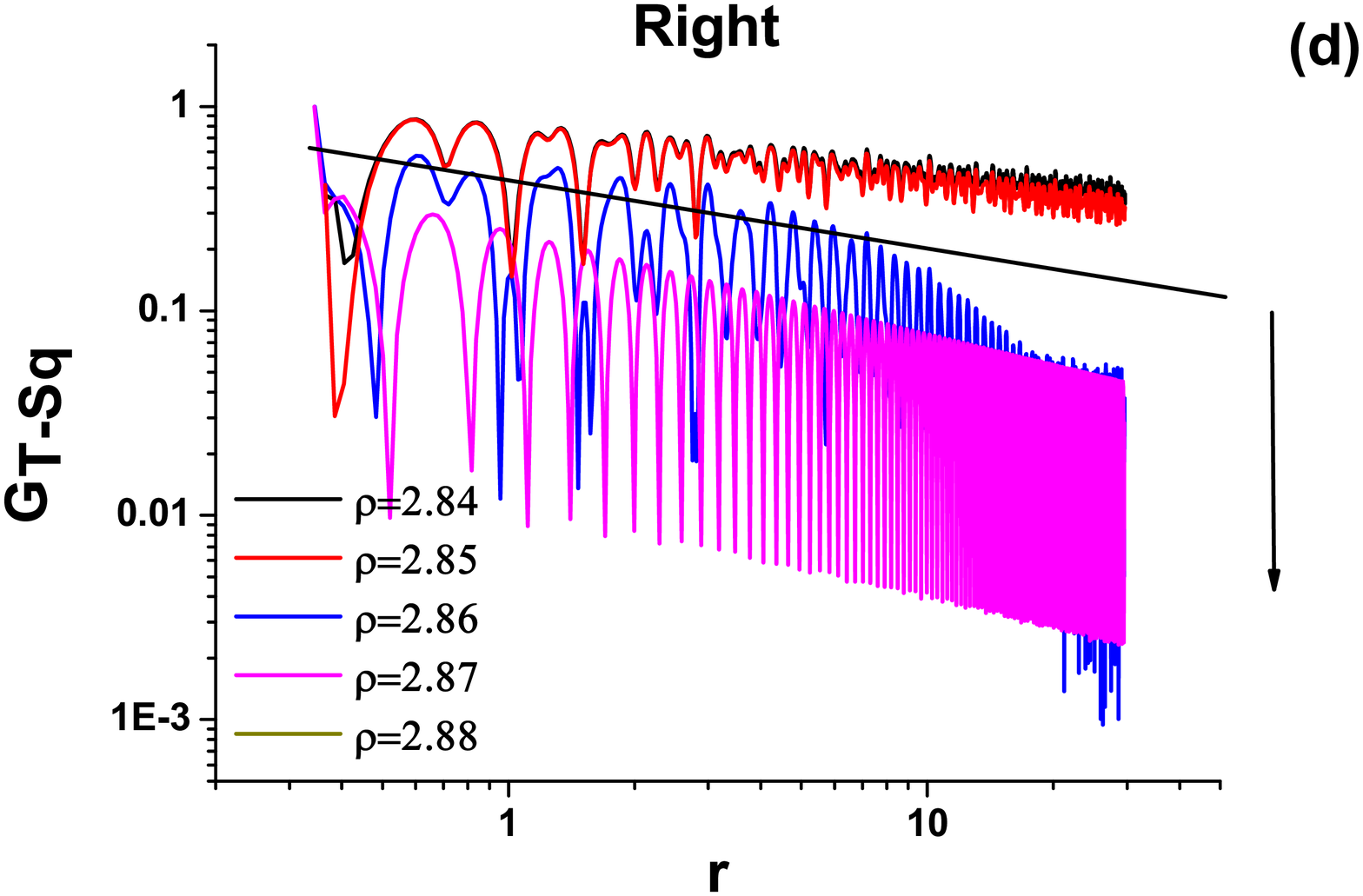}%

\end{center}

\caption{\label{fig:g4gtsq} (a) Orientational correlation function
$G_4$ at the low density branch of the square crystal melting. (b)
Translational correlation function $G_T^{sq}$ in the same region.
(c) Orientational correlation function at the right part of the
melting line of the square crystal. (d) Translational correlation
function in the same region. Arrows at all panels mark the
direction of the density increase.}
\end{figure}

\begin{figure}
\begin{center}
\includegraphics[width=6cm,height=6cm]{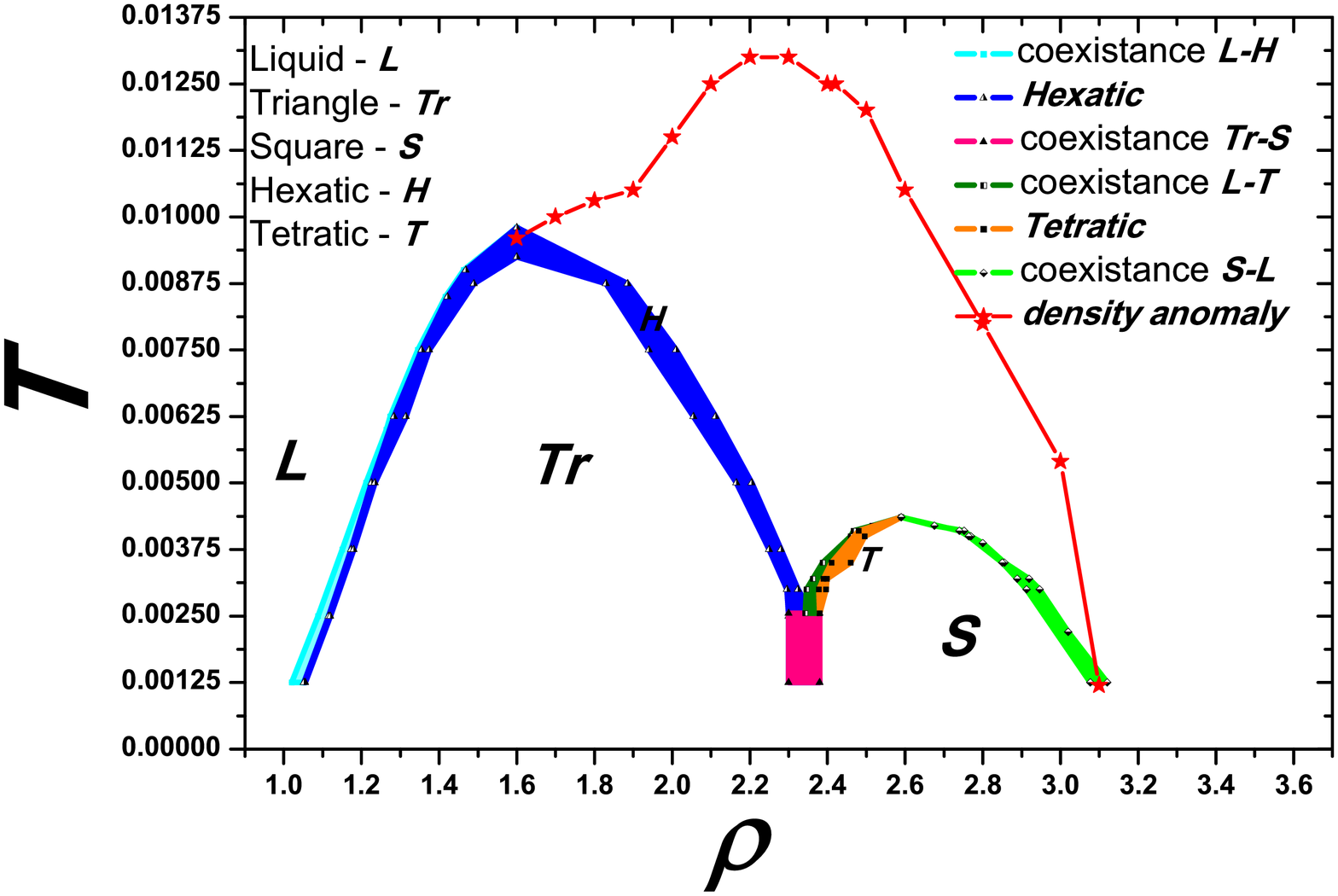}%

\end{center}

\caption{\label{fig:pdtrsq} Low density part of the phase
diagram.}
\end{figure}

Phase transformation of the tetratic phase into the square crystal
which takes place at the left branch is a continuous transition.
The location of this transition is determined by the $G_4$ and
$G_T^{sq}$ correlation functions shown in Figs. \ref{fig:g4gtsq}
(a)-(d). The resulting phase diagram which includes the melting
lines of the triangular and square crystals is shown in Fig.
\ref{fig:pdtrsq}.

\begin{figure}
\begin{center}
\includegraphics[width=6cm,height=6cm]{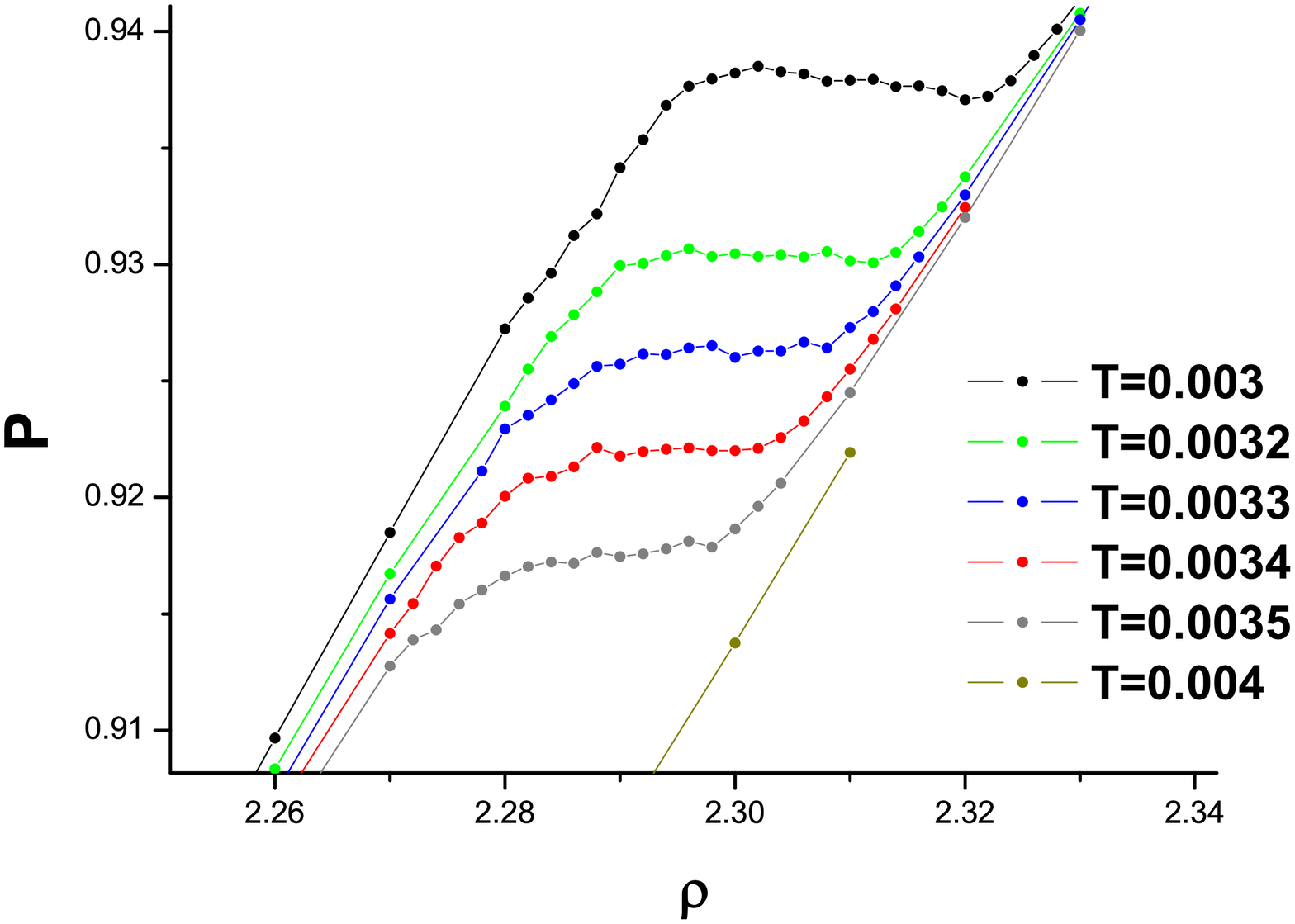}%

\end{center}

\caption{\label{fig:tricr} Equations of state in the vicinity of
the triple point.}
\end{figure}

Special care should be taken in the vicinity of the triple point
of liquid-hexatic-tetratic phases. Fig. \ref{fig:tricr} shows a
set of isotherms in the vicinity of the triple point. One can see
two important features of these equations of state. First of all,
as the temperature increases the curves go down, i.e. the pressure
decreases with temperature. This phenomenon is known as the
density anomaly. We study the region of existence of the density
anomaly and put it on the phase diagram (Fig. \ref{fig:pdtrsq}).
The second is that at low temperatures the eos demonstrates the
Mayer-Wood loop, while at the temperatures above $T_{tcp}=0.0034$
the loop disappears. It leads to a conclusion that the melting
scenario changes at $T_{tcp}$. While above this temperature
melting proceeds in accordance to BKTHNY scenario, below $T_{tcp}$
melting becomes the first order phase transition. The point at
$T_{tcp}=0.0034$ and density $\rho_{tcp}=2.296$ is a tricritical
point of the system, where the BKT continuous hexatic-isotropic
liquid transition  (BKTHNY scenario) is replaced by the
first-order hexatic-isotropic liquid transition (scenario,
proposed in \cite{foh1,foh2,foh3,foh4,foh5,foh6}).

\bigskip
\section{Conclusions}
We present the computer simulation study of a two-dimensional
system of purely repulsive soft colloidal particles interacting
via the Hertz potential in a wide range of densities. This
potential describes the elastic interaction of weakly deformable
bodies and can be a reliable model for qualitative description of
behavior of soft macromolecules, like globular micelles and star
polymers. We find a large number of ordered phases, including the
dodecagonal quasicrystal, and analyze the melting scenarios of low
density triangle and square phases. It is interesting that
depending on the position on the phase diagram the system can melt
both through the first order transition and in accordance with the
BKTHNY scenario and also in accordance with recently proposed
two-stage melting with the first order hexatic-isotropic liquid
transition and continuous BKT type solid-hexatic transition. It is
found that in the liquid phase in the reentering melting region
between the triangle and square lattices there is a water-like
density anomaly. We also demonstrate the possibility of the
tricritical point on the melting line of the triangle phase.

It should be also noted, that the nature of the first-order
liquid-hexatic transition is not completely understood because
conventional theories like the BKTHNY are not capable to describe
the first-order liquid-hexatic transition. However, it is known
that the BKT transition can be made first order by reducing the
core energy $E_c$ of the  corresponding topological defect
(disclination) below some critical value
\cite{fish94,ryz93,ryz94,ryz96}. Taking into account that the
microscopic calculation of $E_c$ is not available now, one can
conclude that the further investigations of the nature of
first-order hexatic-isotropic liquid transition are necessary.

\bigskip

The authors are grateful to Daan Frenkel, S.M.
Stishov, V.V. Brazhkin and E. E. Tareyeva for valuable
discussions. This work was carried out using computing resources
of the federal collective usage center "Complex for simulation and
data processing for mega-science facilities" at NRC "Kurchatov
Institute", http://ckp.nrcki.ru, and supercomputers at Joint
Supercomputer Center of the Russian Academy of Sciences (JSCC
RAS). The work was supported by the Russian Science Foundation
(Grant No 14-22-00093).


\begin{thebibliography}{99}

\bibitem{lan} L. D.Landau, Phys. Z. Sowjetunion {\bf 11}, 26
(1937).

\bibitem{p1} R. E. Peierls, Helv. Phys. Acta {\bf 7}, 81 (1934).

\bibitem{p2} R. E. Peierls, Ann. Inst. Henri Poincar´e {\bf 5}, 177
(1935).


\bibitem{mermin} N. D. Mermin, Phys. Rev. {\bf 176}, 250 (1968).

\bibitem{pu2017} V. N. Ryzhov, E. E. Tareyeva, Yu. D. Fomin, E. N. Tsiok
Physics-Uspekhi {\bf 60}, 857 (2017) (DOI:
10.3367/UFNe.2017.06.038161).

\bibitem{chui83} S.T. Chui, Phys. Rev. B {\bf 28}, 178 (1983).

\bibitem{ryzhovJETP} V.N. Ryzhov, Zh. Eksp. Teor. Fiz. {\bf 100},
1627 (1991). [Sov. Phys. JETP {\bf 73}, 899 (1991)].

\bibitem{RT1} V.N. Ryzhov and E.E. Tareyeva, Physica A {\bf 314},
396 (2002).

\bibitem{rto1} V.N. Ryzhov and E.E. Tareyeva, Phys. Rev. B {\bf 51}
8789 (1995).

\bibitem{rto2}   V.N. Ryzhov and E.E. Tareeva, Zh. Eksp. Teor.
Fiz. {\bf 108}, 2044 (1995) [J. Exp. Theor. Phys. {\bf 81}, 1115
(1995)].

\bibitem{ber1} V. L. Berezinskii, Zhur. Eksp. Teor. Fiz. {\bf 59}, 907 (1970); Sov.
Phys. JETP {\bf 32}, 493 (1970).

\bibitem{ber2} V. L. Berezinskii, Zhur. Eksp. Teor. Fiz. {\bf 61}, 1144 (1971); Sov.
Phys. JETP {\bf 34}, 610 (1971).

\bibitem{kosthoul72} J. M. Kosterlitz and D. J. Thouless, J. Phys. C: Solid St. Phys.
{\bf 5}, L124 (1972).

\bibitem{kosthoul73} J. M. Kosterlitz and D.J. Thouless, J. Phys. C {\bf 6},
1181 (1973).

\bibitem{kost} J. M. Kosterlitz, Rep. Prog. Phys. {\bf 79}, 026001
(2016).

\bibitem{halpnel1} B.I. Halperin and D.R.Nelson, Phys. Rev. Lett. {\bf
41}, 121 (1978).

\bibitem{halpnel2} D.R.Nelson and B.I. Halperin, Phys. Rev. B {\bf 19},
2457 (1979).

\bibitem{halpnel3} A.P. Young,  Phys. Rev. B {\bf 19}, 1855
(1979).


\bibitem{keim1} Urs Gasser, C. Eisenmann, G. Maret, and P. Keim, ChemPhysPhysChem, {\bf 11},
963 (2010).

\bibitem{zanh} K. Zahn and G. Maret, Phys. Rev. Lett. {\bf
85}, 3656 (2000).

\bibitem{keim2} P. Keim, G. Maret, and H.H. von Grunberg,
Phys. Rev. E {\bf 75}, 031402 (2007).

\bibitem{keim3} S. Deutschlander, T. Horn, H. Lowen, G. Maret, and P. Keim,
Phys. Rev. Lett. {\bf 111}, 098301 (2013).

\bibitem{keim4} T. Horn, S. Deutschlander, H. Lowen, G. Maret, and P. Keim,
Phys. Rev. E, {\bf 88}, 062305 (2013).

\bibitem{foh1} E.P. Bernard and W. Krauth, Phys. Rev. Lett. {\bf 107},
155704 (2011).

\bibitem{foh2} M. Engel, J.A. Anderson, S.C. Glotzer, M. Isobe, E.P. Bernard,
W. Krauth, Phys. Rev. E {\bf 87}, 042134 (2013).

\bibitem{foh3} W. Qi, A. P. Gantapara and M. Dijkstra, Soft Matter {\bf 10},
5449 (2014).

\bibitem{foh4} S.C. Kapfer and W. Krauth, Phys. Rev. Lett. {\bf 114},
035702 (2015).

\bibitem{foh5} W.K. Qi, S.M. Qin, X.Y. Zhao, and
Yong Chen, J. Phys.: Condens. Matter {\bf 20}, 245102 (2008).

\bibitem{foh6} W. Qi and M. Dijkstra, Soft Matter {\bf 11},  2852 (2015) (DOI:
10.1039/c4sm02876g).

\bibitem{si} G. Algara-Siller, O. Lehtinen, F. C. Wang, R. R. Nair, U. Kaiser, H. A.Wu,
A. K. Geim, I. V. Grigorieva, Nature  {\bf 519}, 443 (2015).

\bibitem{siron} Jiong Zhao et al.,  Science {\bf 343}, 1228 (2014).

\bibitem{str1} Ronen Zangi and Alan E. Mark, Phys. Rev. Lett. {\bf 91}, 025502
(2003).

\bibitem{str2} P. Kumar, S. V. Buldyrev, F. W. Starr, N. Giovambattista, and H.
Eugene Stanley, Phys. Rev. E {\bf 72}, 051503 (2005).

\bibitem{str3} Sungho Han, M. Y. Choi, Pradeep Kumar and H. Eugene Stanley, Nature Physics {\bf 6}, 685
(2010).

\bibitem{str4} Jessica C. Johnston, Noah Kastelowitz, and Valeria Molinero, J. Chem. Phys. {\bf 133}, 154516 (2010).

\bibitem{str5} Hamid Mosaddeghi, Saman Alavi, M. H. Kowsari, and Bijan Najafi, J. Chem. Phys. {\bf 137}, 184703 (2012).

\bibitem{str6} Ahmad M. Almudallal, Sergey V. Buldyrev, and Ivan
Saika-Voivod, J. Chem. Phys. {\bf 137}, 034507 (2012).

\bibitem{str7} T. Dotera, T. Oshiro, P. Ziherl, Nature {\bf 509}, 208
(2014).

\bibitem{str8} H. Pattabhiramana and M. Dijkstra, J. Chem. Phys. {\bf 146}, 114901
(2017).

\bibitem{str9} Q. Meng, C. N. Varney, H. Fangohr, and E. Babaev , Phys. Rev. B
{\bf 90}, 020509(R) (2014).

\bibitem{dfrt1} D.E. Dudalov, Yu.D. Fomin, E.N. Tsiok, and V.N. Ryzhov,
J. Phys.: Conference Series {\bf 510}, 012016 (2014)
(doi:10.1088/1742-6596/510/1/012016).

\bibitem{dfrt2} D.E. Dudalov, Yu.D. Fomin, E.N. Tsiok, and V.N.
Ryzhov, J. Chem. Phys. {\bf 141}, 18C522 (2014).

\bibitem{dfrt3} D.E. Dudalov, Yu.D. Fomin, E.N. Tsiok, and V.N.
Ryzhov, Soft Matter {\bf 10}, 4966 (2014).

\bibitem{dfrt4} E.S. Chumakov, Y.D. Fomin, E.L. Shangina, E.E. Tareyeva, E.N.
Tsiok, V.N. Ryzhov, Physica A {\bf 432}, 279 (2015).

\bibitem{dfrt5}  E. N. Tsiok, D. E. Dudalov, Yu. D. Fomin, and V. N.
Ryzhov,  Phys. Rev. E {\bf 92}, 032110 (2015).

\bibitem{dfrt6} E. N. Tsiok, Y. D. Fomin, V. N. Ryzhov, Physica A {\bf 490}, 819 (2018).

\bibitem{dfrt7} N.P. Kryuchkov, S.O. Yurchenko, Yu. D. Fomin, E. N.
Tsiok, and V. N. Ryzhov, arXiv:1712.04707.

\bibitem{trus} William D. Piñeros, Michael Baldea, and Thomas M.
Truskett, J. Chem. Phys. {\bf 145}, 054901 (2016).

\bibitem{jcp2008} Y.D. Fomin, N.V. Gribova, V.N. Ryzhov,
S.M Stishov, and D. Frenkel, J. Chem. Phys. \textbf{129}, 064512
(2008).

\bibitem{wepre} N.V. Gribova, Y.D. Fomin, D. Frenkel, and
V.N. Ryzhov, Phys. Rev. E \textbf{79} 051202 (2009).

\bibitem{we_inv} Y.D. Fomin, E.N. Tsiok, and V.N. Ryzhov, J.
Chem. Phys. {\bf 135}, 234502 (2011).


\bibitem{we2011} Y.D. Fomin, E.N. Tsiok, and V.N. Ryzhov,
J. Chem. Phys. {\bf 135}, 124512 (2011).


\bibitem{RCR} R.E. Ryltsev, N.M. Chtchelkatchev, and V.N. Ryzhov, Phys. Rev. Lett. {\bf 110},
025701 (2013).

\bibitem{we2013-2} Y.D. Fomin, E.N. Tsiok, and V.N. Ryzhov,
Phys. Rev. E {\bf 87}, 042122 (2013).

\bibitem{buld2009} S.V. Buldyrev, G. Malescio, C.A. Angell, N. Giovambattista,
S. Prestipino, F. Saija, H.E. Stanley, and L. Xu, J. Phys.:
Condens. Matter {\bf 21},  504106 (2009).

\bibitem{fr1} P. Vilaseca and G. Franzese, Journal of Non-Crystalline
Solids {\bf 357},  419 (2011).

\bibitem{fr2} G. Franzese, J. Mol. Liq. {\bf 136}, 267 (2007).

\bibitem{fr3} Pol Vilaseca and Giancarlo Franzese, J. Chem. Phys. {\ bf 133},
084507 (2010).

\bibitem{fr4} F. Leonia and G. Franzese, J.Chem.Phys. {\bf 141},
174501 (2014).

\bibitem{barbosa} A. B. de Oliveira, P. A. Netz, T. Colla, and M. C. Barbosa, J.
Chem. Phys. {\bf 124}, 084505 (2006).


\bibitem{barbosa1} L.B. Krott and M.C. Barbosa, J. Chem. Phys. {\bf 138}
084505 (2013).

\bibitem{barbosa2} L. B. Krott and M. C. Barbosa, Phys. Rev. E {\bf 89},
012110 (2014).

\bibitem{barbosa3} L. B. Krott, J. R. Bordin, and M.C. Barbosa, J. Phys. Chem. B {\bf 119},
291 (2015).

\bibitem{buld2d} A.M. Almudallal, S.V. Buldyrev, and I. Saika-Voivod, J. Chem. Phys. {\bf 137}
034507 (2012).

\bibitem{scala} M.R. Sadr-Lahijany, A. Scala, S.V. Buldyrev, H.E. Stanley, Phys.
Rev. Lett. {\bf 81}, 4895 (1998).

\bibitem{prest2} S. Prestipino, F. Saija, and P.V. Giaquinta, J. Chem.
Phys. {\bf 137}, 104503 (2012).

\bibitem{prest1} S. Prestipino, F. Saija, and P.V. Giaquinta, Phys. Rev. Lett. {\bf 106},
235701 (2011).

\bibitem{lanl} L. D. Landau and E. M. Lifshitz, Theory of Elasticity, 3rd ed.
(Pergamon, New York, 1986).


\bibitem{hertz3d} J. C. Pamies, A. Cacciuto, and D. Frenkel, J.
Chem. Phys. 131 , 044514 (2009).

\bibitem{rosbreak} Yu. D. Fomin, V. N. Ryzhov, N. V. Gribova,
Phys. Rev. E 81, 061201 (2010).

\bibitem{strcriteria} Yu. D. Fomin, V. N. Ryzhov, B. A. Klumov, and E. N.
Tsiok, J. Chem. Phys. 141, 034508 (2014).


\bibitem{ros} Ya. Rosenfeld, Phys. Rev. A 15, 2545 (1977).

\bibitem{ros1} Ya. Rosenfeld, J. Phys.: Condens. Matter 11, 5415
(1999).

\bibitem{miller} Miller W.L., Cacciuto A. Two-dimensional packing of soft
particles and the soft generalized thomson problem. Soft Matter 7,
7552 (2011).

\bibitem{hertzqc} M. Zu, P. Tan and N. Xu, Nature Comm. 8, 2089
(2017).

\bibitem{hertzmelt} M. Zum J. Liu, H. Tong and N. Xu, Phys. Rev.
Lett. 117, 085702 (2016).


\bibitem{book_fs} Daan Frenkel and Berend Smit, {\it Understanding
molecular simulation (From Algorithms to Applications), 2nd
Edition} (Academic Press, 2002).


\bibitem{lammps} http://lammps.sandia.gov/

\bibitem{fish94} Y. Levin, X. Lee, M. E. Fisher, Phys. Rev. Lett. {\bf 93}, 2716
(1994).

\bibitem{ryz93} V. N. Ryzhov, E. E. Tareyeva, Phys. Rev. B {\bf 48}, 12907
(1993).

\bibitem{ryz94} V. N. Ryzhov, E. E. Tareyeva, Phys. Rev. B {\bf 49}, 6162
(1994).

\bibitem{ryz96} D. Y. Irz, V. N. Ryzhov, E. E. Tareyeva, Phys. Rev. B {\bf 54}, 3051
(1996).




\end{thebibliography}
\end{document}